\documentclass[apjl]{emulateapj}

\usepackage{epsfig}
\usepackage{xspace}
\usepackage{amsmath}
\usepackage{color}

\newcommand{\mpch}{\>h^{-1}{\rm {Mpc}}}

\newcommand{\msunh}{\>h^{-1}\rm M_\odot}

\def\gcm3{\mathrm{g} / \mathrm{cm}^3}

\def\gtsima{$\; \buildrel > \over \sim \;$}
\def\ltsima{$\; \buildrel < \over \sim \;$}
\def\prosima{$\; \buildrel \propto \over \sim \;$}
\def\gsim{\lower.7ex\hbox{\gtsima}}
\def\lsim{\lower.7ex\hbox{\ltsima}}
\def\simgt{\lower.7ex\hbox{\gtsima}}
\def\simlt{\lower.7ex\hbox{\ltsima}}
\def\simpr{\lower.7ex\hbox{\prosima}}

\slugcomment{Submitted to ApJ}

\shorttitle{WL Mass Calibration of ACTPol Clusters with HSC}
\shortauthors{Miyatake, Battaglia, Hilton et al.}

\begin{document}

\title{Weak-Lensing Mass Calibration of ACTPol Sunyaev-Zel'dovich Clusters with the Hyper Suprime-Cam Survey}
\author{Hironao Miyatake\altaffilmark{1,2,3,4}, 
Nicholas Battaglia\altaffilmark{5,6,7}, 
Matt Hilton\altaffilmark{8}, 
Elinor Medezinski\altaffilmark{7}, 
Atsushi J. Nishizawa\altaffilmark{1,2},
Surhud More\altaffilmark{4},
Simone Aiola\altaffilmark{9},
Neta Bahcall\altaffilmark{7},
J. Richard Bond\altaffilmark{10},
Erminia Calabrese\altaffilmark{11},
Steve K. Choi\altaffilmark{9},
Mark J. Devlin\altaffilmark{12},
Joanna Dunkley\altaffilmark{7,9},
Rolando Dunner\altaffilmark{13},
Brittany Fuzia\altaffilmark{14},
Patricio Gallardo\altaffilmark{15},
Megan Gralla\altaffilmark{16},
Matthew Hasselfield\altaffilmark{17,18},
Mark Halpern\altaffilmark{19},
Chiaki Hikage\altaffilmark{4},
J. Colin Hill\altaffilmark{20,5},
Adam D. Hincks\altaffilmark{21},
Ren\'ee Hlo\v{z}ek\altaffilmark{22},
Kevin Huffenberger\altaffilmark{14},
John P. Hughes\altaffilmark{23,5},
Brian Koopman\altaffilmark{15},
Arthur Kosowsky\altaffilmark{24},
Thibaut Louis\altaffilmark{25},
Mathew S. Madhavacheril\altaffilmark{7},
Jeff McMahon\altaffilmark{26},
Rachel Mandelbaum\altaffilmark{27},
Tobias A. Marriage\altaffilmark{28},
Lo\"ic Maurin\altaffilmark{13},
Satoshi Miyazaki\altaffilmark{29},
Kavilan Moodley\altaffilmark{8},
Ryoma Murata\altaffilmark{4,30},
Sigurd Naess\altaffilmark{5},
Laura Newburgh\altaffilmark{31},
Michael D. Niemack\altaffilmark{15},
Takahiro Nishimichi\altaffilmark{4},
Nobuhiro Okabe\altaffilmark{32,33},
Masamune Oguri\altaffilmark{34,30,4},
Ken Osato\altaffilmark{30},
Lyman Page\altaffilmark{9},
Bruce Partridge\altaffilmark{35},
Naomi Robertson\altaffilmark{36},
Neelima Sehgal\altaffilmark{37},
Blake Sherwin\altaffilmark{38,39},
Masato Shirasaki\altaffilmark{29},
Jonathan Sievers\altaffilmark{40},
Crist\'obal Sif\'on\altaffilmark{7},
Sara Simon\altaffilmark{26},
David N. Spergel\altaffilmark{5,7},
Suzanne T. Staggs\altaffilmark{9},
George Stein\altaffilmark{41,10},
Masahiro Takada\altaffilmark{4},
Hy Trac\altaffilmark{27},
Keiichi Umetsu\altaffilmark{42},
Alex van Engelen\altaffilmark{10},
Edward J. Wollack\altaffilmark{43}}
\altaffiltext{1}{Institute for Advanced Research, Nagoya University, Nagoya 464-8601, Japan}
\altaffiltext{2}{Division of Physics and Astrophysical Science, Graduate School of Science, Nagoya University, Nagoya 464-8602, Japan}
\altaffiltext{3}{Jet Propulsion Laboratory, California Institute of Technology, Pasadena, CA 91109, USA}
\altaffiltext{4}{Kavli Institute for the Physics and Mathematics of the Universe (Kavli IPMU, WPI),
UTIAS, The University of Tokyo, Chiba 277-8583, Japan}
\altaffiltext{5}{Center for Computational Astrophysics, Flatiron Institute, New York, NY 10010, USA}
\altaffiltext{6}{Department of Astronomy, Cornell University, Ithaca, NY 14853, USA}
\altaffiltext{7}{Department of Astrophysical Sciences, Princeton University, Princeton,
  NJ 08544, USA}
\altaffiltext{8}{Astrophysics \& Cosmology Research Unit, School of Mathematics, Statistics \& Computer Science, University of KwaZulu-Natal, Westville Campus, Durban 4041, South Africa}
\altaffiltext{9}{Department of Physics, Princeton University, Princeton, NJ 08544, USA}
\altaffiltext{10}{Canadian Institute for Theoretical Astrophysics, University of Toronto, Toronto, ON, M5S 3H8, Canada}
\altaffiltext{11}{School of Physics and Astronomy, Cardiff University, The Parade, Cardiff, CF24 3AA, UK}
\altaffiltext{12}{Department of Physics and Astronomy, University of Pennsylvania, Philadelphia, PA 19104, USA}
\altaffiltext{13}{Instituto de Astrof\'isica and Centro de Astro-Ingenier\'ia, Facultad de F\'isica, Pontificia Universidad Cat\'olica de Chile, Av. Vicu\~na Mackenna 4860, 7820436 Macul, Santiago, Chile}
\altaffiltext{14}{Department of Physics, Florida State University, Tallahassee, Florida 32306}
\altaffiltext{15}{Department of Physics, Cornell University, Ithaca, NY 14853, USA}
\altaffiltext{16}{Department of Astronomy/Steward Observatory, University of Arizona, 933 N. Cherry Ave., Tucson, AZ 85721}
\altaffiltext{17}{Institute for Gravitation and the Cosmos, The Pennsylvania State University, University Park, PA 16802, USA}
\altaffiltext{18}{Department of Astronomy and Astrophysics, The Pennsylvania State University, University Park, PA 16802, USA}
\altaffiltext{19}{Department of Physics and Astronomy, University of British Columbia, Vancouver, BC, Canada V6T 1Z4}
\altaffiltext{20}{School of Natural Sciences, Institute for Advanced Study, Princeton, NJ 08540, USA}
\altaffiltext{21}{Department of Physics, University of Rome ``La Sapienza", I-00185 Rome, Italy}
\altaffiltext{22}{Dunlap Institute, University of Toronto, Toronto, ON, Canada M5S 3H4}
\altaffiltext{23}{Department of Physics and Astronomy, Rutgers University, Piscataway, NJ 08854, USA}
\altaffiltext{24}{Department of Physics and Astronomy, University of Pittsburgh, Pittsburgh, PA 15260 ,USA}
\altaffiltext{25}{Laboratoire de l’Acc\'el\'erateur Lin\'eaire, Universit\'e Paris-Sud, CNRS/IN2P3, Universit\'e Paris-Saclay, Orsay, France}
\altaffiltext{26}{Department of Physics, University of Michigan, Ann Arbor, MI 48103, USA}
\altaffiltext{27}{McWilliams Center for Cosmology, Department of Physics, Carnegie Mellon University, Pittsburgh, PA 15213, USA}
\altaffiltext{28}{Dept. of Physics and Astronomy, Johns Hopkins University, Baltimore, MD 21218, USA}

\altaffiltext{29}{National Astronomical Observatory of Japan, Tokyo 181-8588, Japan}
\altaffiltext{30}{Department of Physics, University of Tokyo, Tokyo, 113-0033, Japan}
\altaffiltext{31}{Department of Physics, Yale University, New Haven, CT 06520}
\altaffiltext{32}{Department of Physical Science, Hiroshima University, Hiroshima 739-8526, Japan}
\altaffiltext{33}{Hiroshima Astrophysical Science Center, Hiroshima University, Hiroshima, 739-8526, Japan}
\altaffiltext{34}{Research Center for the Early Universe, University of Tokyo, Tokyo 113-0033, Japan}
\altaffiltext{35}{Department of Physics and Astronomy, Haverford College, Haverford, PA 19041, USA}
\altaffiltext{36}{Department of Astrophysics, University of Oxford, Keble Road, Oxford OX1 3RH, UK}
\altaffiltext{37}{Physics and Astronomy Department, Stony Brook University, Stony Brook, NY 11794}
\altaffiltext{38}{Department of Applied Mathematics and Theoretical Physics, University of Cambridge, Wilberforce Road, Cambridge CB3 0WA, UK}
\altaffiltext{39}{Kavli Institute for Cosmology, University of Cambridge, Madingley Road, Cambridge CB3 OHA, UK}
\altaffiltext{40}{Astrophysics \& Cosmology Research Unit, School of School of Chemistry \& Physics, University of KwaZulu-Natal, Westville Campus, Durban 4041, South Africa}
\altaffiltext{41}{Department of Astronomy and Astrophysics, University of Toronto, Toronto, ON, M5S 3H4, Canada}
\altaffiltext{42}{Institute of Astronomy and Astrophysics, Academia Sinica, Taipei 10617, Taiwan}
\altaffiltext{43}{NASA/Goddard Space Flight Center, Greenbelt, MD 20771, USA}

\email{hironao.miyatake@iar.nagoya-u.ac.jp}

\begin{abstract}
We present weak-lensing measurements using the first-year data from the Hyper Suprime-Cam Strategic Survey Program on the Subaru telescope for eight galaxy clusters selected through their thermal Sunyaev-Zel'dovich (SZ) signal measured at 148 GHz with the Atacama Cosmology Telescope Polarimeter experiment. The overlap between the two surveys in this work is 33.8 square degrees, before masking bright stars. The signal-to-noise ratio of individual cluster lensing measurements ranges from 2.2 to 8.7, with a total of 11.1 for the stacked cluster weak-lensing signal. We fit for an average weak-lensing mass distribution using three different profiles, a Navarro-Frenk-White profile, a dark-matter-only emulated profile, and a full cosmological hydrodynamic emulated profile. We interpret the differences among the masses inferred by these models as a systematic error of 10\%, which is currently smaller than the statistical error. We obtain the ratio of the SZ-estimated mass to the lensing-estimated mass (the so-called hydrostatic mass bias $1-b$) of $0.74^{+0.13}_{-0.12}$, which is comparable to previous SZ-selected clusters from the Atacama Cosmology Telescope and from the {\sl Planck} Satellite. We conclude with a discussion of the implications for cosmological parameters inferred from cluster abundances compared to cosmic microwave background primary anisotropy measurements.
\end{abstract}


\section{Introduction}

Measurements of the abundance of galaxy clusters can be used as a probe of the growth of structure in the Universe. In particular, since clusters are the rarest and most massive collapsed structures, their abundances as a function of mass and redshift are particularly sensitive to the normalization of the matter power spectrum, $\sigma_8$, and the matter density, $\Omega_{\rm m}$ \citep[][for review]{Voit2005,AEM2011}. However, current cluster abundance measurements are limited by systematic uncertainties in observable-to-mass relations, as reported in recent measurements  \citep[e.g.,][]{Vikhlinin2009,Rozo2010, Vand2010,Sehgal2011,Benson2013,Hass2013,Planckcounts,Mantz2014,PlnkSZCos2015,Mantz2015,deHaan2016}. Thus, accurate empirical calibrations of observable-mass relations are essential for cluster surveys to fully reach their potential.

Samples of clusters are assembled based on several observables such as the density and concentration of galaxies in optical/IR observations \citep[e.g.,][]{Rykoff:2014}, the projected density map measured by weak lensing \citep[e.g.,][]{Miyazaki:2018b}, the X-ray emission from cluster hot gas \citep[e.g.,][]{Pacaud:2016}, and the thermal Sunyaev Zel'dovich (SZ, and hereafter SZ refers to the thermal SZ) effect \citep{SZ1969,SZ1972}, a characteristic spectral distortion in the cosmic microwave background (CMB) induced by inverse Compton scattering between CMB photons and hot ionized electrons. The spectral shape of the SZ effect is a decrement in thermodynamic temperature at frequencies below 217 GHz and excess at higher frequencies, and its amplitude scales with the Compton-$y$ parameter. Among these observables, the SZ effect is unique because the detection efficiency is nearly independent of redshift as long as the beam size is about arcminute scale. As a consequence, SZ-selected cluster samples have well-behaved selection functions that make it straightforward to calibrate observable-to-mass relations and constrain cosmological parameters. Additionally the integrated SZ signal is a low-scatter proxy for mass \citep[e.g.,][]{Motl2005,Nagai2006,Stanek2010,BBPS1,Sembolini2013} that is fairly robust against cluster astrophysics \citep[e.g.,][]{Nagai2006,BBPS1,Planelles2017}

Current and recent CMB experiments like the Atacama Cosmology Telescope \citep[ACT;][]{ACTPol}, the South Pole Telescope \citep[SPT;][]{Carlstrom2011}, and the {\sl Planck} satellite \citep{PlanckSum2016} have provided large catalogs of SZ-selected clusters \citep[e.g.,][]{Stan2009,Marriage2011,Reic2013,Hass2013,PlanckClustCat,Bleem2015,PlnkSrc2015,Hilton2017}. In these experiments, different observational techniques are used to measure the integrated Compton-$y$ signal and different SZ-mass scaling relations are used to infer cluster masses. For example, the {\sl Planck} collaboration relies on X-ray observables for their  initial calibrations of the Compton-$y$ to mass relation \citep{PlanckClustCat}. However, the determination of a cluster's total mass (including dark matter) from X-ray observables assumes that the intracluster medium is in hydrostatic equilibrium. Such physical assumptions can be a source of systematic uncertainties in cluster mass estimates \citep[e.g.,][]{Evrard1990,Rasia2004,lau2009,BBPS1,Nelson2012,Rasia2012}.

The technique of weak lensing offers direct measurement of the total matter distribution of a galaxy cluster (baryonic and dark matter), and can thus provide an unbiased mass calibration. Weak lensing manifests itself as small but coherent distortions of distant galaxies that result from the gravitational deflection of light due to foreground structures \citep[e.g.,][]{Kaiser1992}. Cluster weak lensing appears as tangential shear of background galaxy shapes around a cluster. Numerous attempts to calibrate SZ masses have been made in the literature using ACT clusters \citep{Miyatake2013,Jee2014,Battaglia2016}, SPT clusters \citep{McI2009,High2012,Schrabback2016,Dietrich2017,Stern2018}, {\sl Planck} clusters \citep{WtG2014,CCCP2015,Penna2017,Sereno2017,Medez2017}, {\sl Planck} and SPT clusters \citep{Gruen2014}, and other massive cluster samples \citep{Marrone2009,Hoek2012,Marrone2012,Smith2016}. The mass calibration is often parametrized as
\begin{equation}
1-b = \frac{M_{\rm SZ}}{M_{\rm true}},
\end{equation}
where $M_{\rm SZ}$ is the SZ mass and $M_{\rm true}$ is the true cluster mass, which for this paper we take to be the weak-lensing mass $M_{\rm WL}$. This ratio can be taken for individual clusters or for an ensemble average and these values will be consistent as long as the appropriate weights are used \citep{Medez2017}. Recently, \cite{PlnkSZCos2015}
reported a disagreement between $1-b$ obtained by weak-lensing calibrations of {\sl Planck} SZ cluster masses \citep[e.g.,][]{WtG2014,CCCP2015} and that inferred from reconciling the {\sl Planck} primary CMB parameters with the {\sl Planck} SZ cluster counts. This disagreement is not statistically significant ($\sim 2\sigma$) and will decrease after accounting for additional bias corrections, like Eddington bias \citep{Battaglia2016} and new optical depth measurements \citep{PlnkTau2016}. However, if such a disagreement persists as the precision of cluster measurements improves, then it could reveal the need for extensions to the standard cosmological model \citep{PlanckCMB2015}, like a non-minimal sum of neutrino masses \citep[e.g.,][]{Wang2005,Shimon2011,Carbone2012,Mak2013,LA2017,MBM2017}, or illuminate additional systematic effects in cluster abundance measurements.

In this paper, we present weak-lensing mass calibrations of SZ-selected clusters. We perform weak-lensing measurements using Subaru Hyper Suprime-Cam (HSC) Strategic Survey Program (SSP) data \citep{Aihara:2018a}. The SZ cluster sample is based on the ACT Polarimeter (ACTPol) two-season cluster catalog \citep{Hilton2017}. Section~\ref{sec:data} describes the details of the ACTPol data and HSC data used in these measurements. Section~\ref{sec:weak_lensing} describes the details of the weak-lensing measurements, including our investigation of systematics. Section~\ref{sec:results} presents the mass calibration of SZ clusters, and we discuss our results and conclude in Section~\ref{sec:conclusion}. Throughout the paper, we adopt the flat-$\Lambda$CDM cosmology with $\Omega_m=0.3$ and $h=0.7$. The SZ masses are quoted in $M_{\rm 500c}$ where the mass enclosed in $R_{\rm 500c}$ is 500 times the critical density of the Universe at the redshift of the cluster. Some of the weak-lensing masses are defined by $M_{\rm 200m}$ where the mass enclosed in $R_{\rm 200m}$ is 200 times the mean matter density. When this is true we converted $M_{\rm 200m}$ to $M_{\rm 500c}$ to compare to the SZ mass.

\section{Data}
\label{sec:data}
\subsection{ACTPol Clusters}
\label{sec:act_data}
The cluster sample used in this work is drawn from the ACTPol two-season cluster catalog \citep{Hilton2017}. The sample was extracted from 148\,GHz observations of a 987.5\,deg$^{2}$ equatorial field that combined data obtained using the original ACT receiver \citep[MBAC;][]{ACT} with the first two seasons of ACTPol data. Details of the ACTPol observations and map making can be found in \citet{Naess2014} and \citet{Louis2017}. Cluster candidates were detected by applying a spatial matched filter to the map, using the Universal Pressure Profile \citep[UPP;][]{Arnaud2010} and its associated SZ signal-mass scaling relation to model the cluster signal. Candidates were then confirmed as clusters and their redshifts measured using optical/IR data, principally from the Sloan Digital Sky Survey \citep[SDSS DR13;][]{Albareti2016}. Cluster masses were estimated by applying the Profile Based Amplitude Analysis technique, introduced in \citet{Hass2013}. In this paper, we use $M_{\rm SZ}$ to refer to SZ-based mass estimates that correspond with $M^{\rm UPP}_{\rm 500c}$ as tabulated in \citet{Hilton2017}. The full cluster sample in \citet{Hilton2017} are all signal-to-noise $(S/N) > 4$ with mass range of roughly $2\times 10^{14}\,$M$_{\odot} <M_{\rm SZ}< 9\times 10^{14}\,$M$_{\odot} $ with a median mass of $M_{\rm SZ} = 3.1\times 10^{14}\,$M$_{\odot}$ and redshift range of roughly $0.15 <z< 1.4$ with a median redshift of $z = 0.49$.

\subsection{HSC - ACT Survey Overlap}

Among the HSC first-year data \citep{Aihara:2018b}, the XMM field in the HSC Wide Layer overlaps with the deepest region of the ACTPol maps - the D6 field at 02$^{\rm h}$30$^{\rm m}$ RA \citep{Naess2014, Louis2017}. The sample studied in this work consists of ACTPol clusters in this region that were detected with SNR$_{\rm 2.4} > 5$ in the \citet{Hilton2017} catalog, where SNR$_{\rm 2.4}$ refers to the signal-to-noise ratio of the cluster, as measured in a map filtered at an angular scale of 2.4$\arcmin$ (refer to Sections~2.2 and 2.3 of \citealt{Hilton2017} for details). Above this threshold, the cluster sample in the HSC S16A region is 100\% pure with complete redshift follow-up. 

\begin{figure}
    \includegraphics[width=\columnwidth]{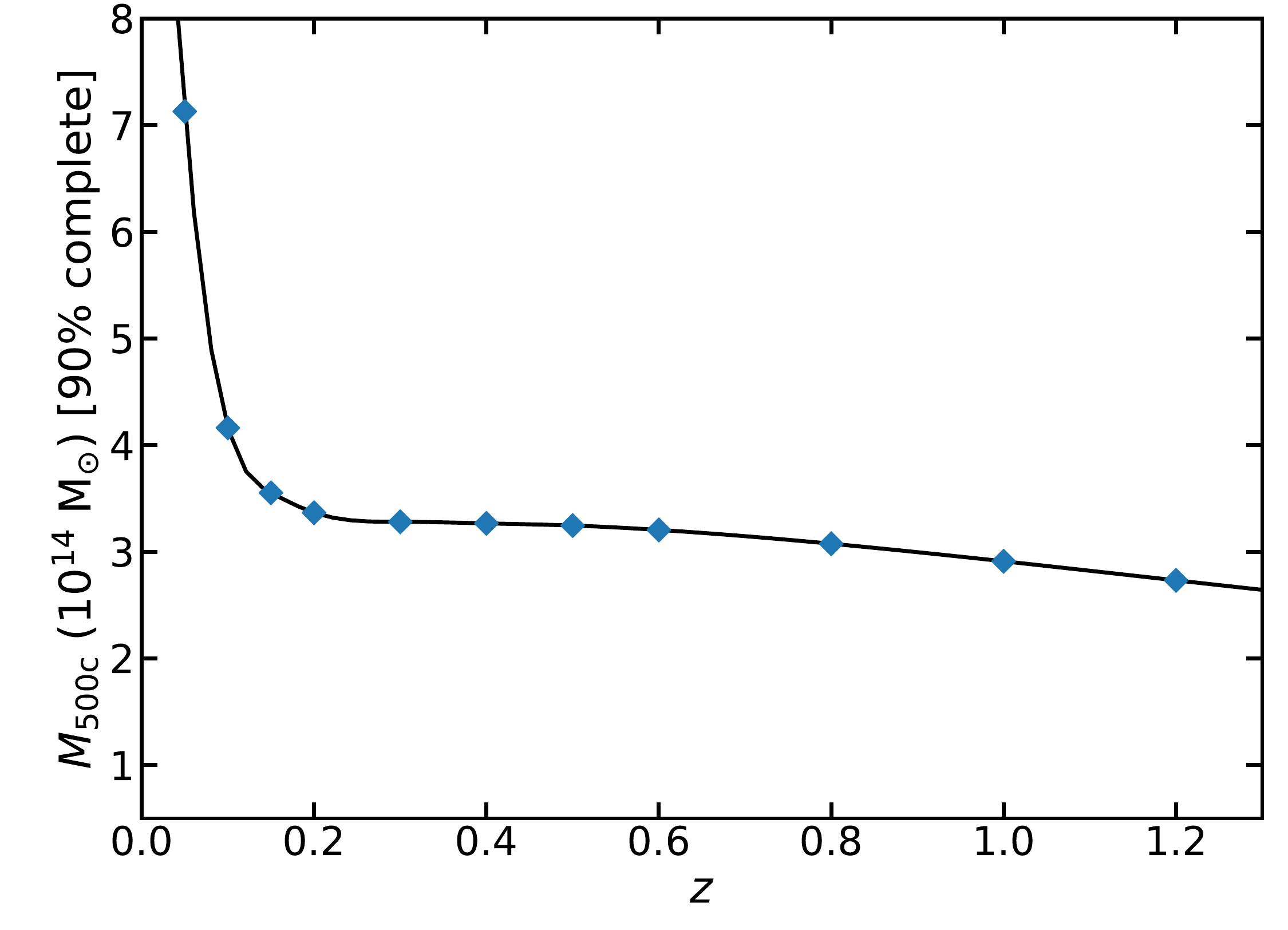}
\caption{Survey-averaged 90\% $M_{\rm SZ}$ completeness limit as a function of redshift, using the UPP and \citet{Arnaud2010} mass scaling relation to model the SZ signal, estimated in the HSC S16A region. The blue diamonds mark the redshifts at which the limit was estimated, and the solid line is a cubic spline fit. In the redshift range $0.2 < z < 1.0$, the average 90\% completeness limit is $M_{\rm SZ} > 3.2 \times 10^{14}$\,M$_\odot$ for SNR$_{2.4} > 5$.}
\label{fig:HSCCompleteness}
\end{figure}

Figure~\ref{fig:HSCCompleteness} shows the 90\% mass completeness limit as a function of redshift across the HSC S16A region, using the UPP model and associated scaling relation to convert the SZ signal into mass, as described in Section 2.4 of \citet{Hilton2017}. Averaged over $0.2 < z < 1$, the sample is 90\% complete for $M_{\rm SZ} > 3.2 \times 10^{14}$\,M$_{\odot}$. This is significantly lower than the equivalent limit of $M_{\rm SZ} > 4.5 \times 10^{14}$\,M$_{\odot}$ obtained when averaging over the whole 987.5\,deg$^{2}$ ACTPol field, since the overlapping HSC S16A region lies in a low noise region of the ACTPol survey D6 \citep[see][]{Naess2014}.

\begin{figure*}
  \begin{center}
    \includegraphics[clip,width=\textwidth]{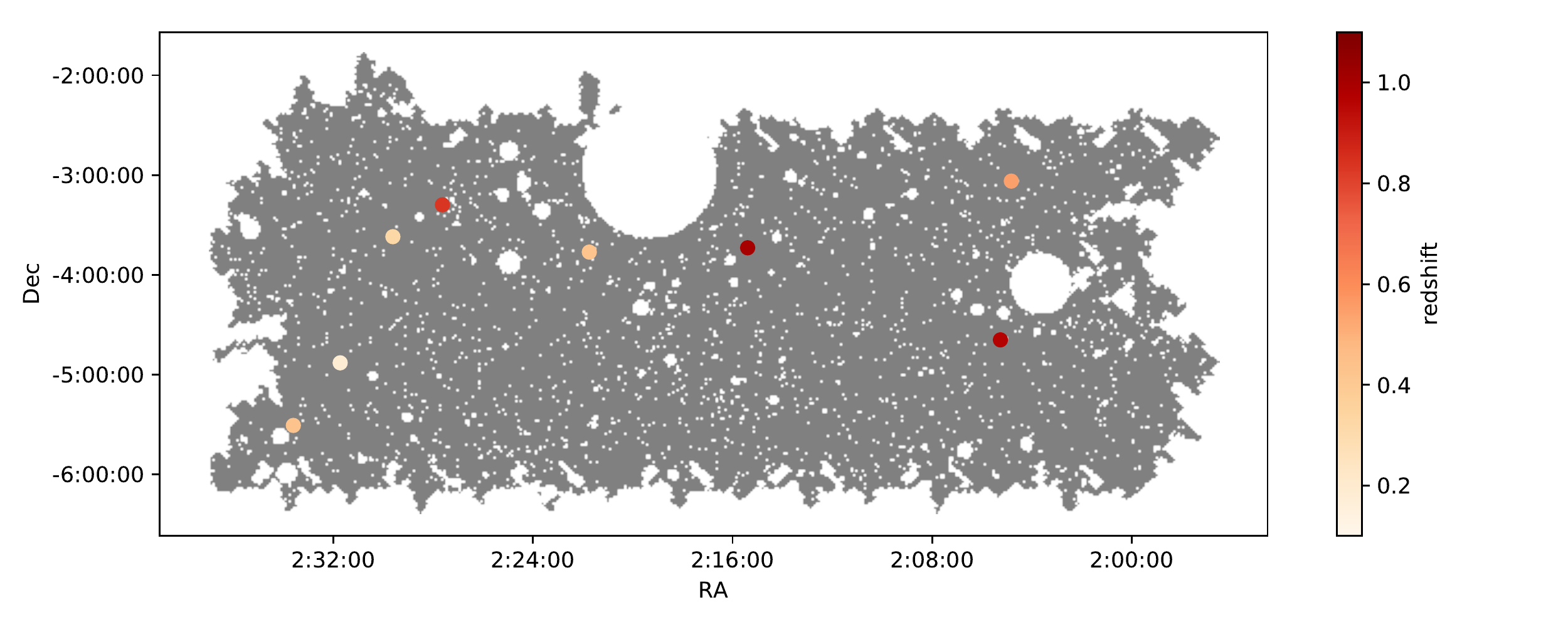}
  \end{center}
\caption{ACTPol SZ-selected clusters in the HSC XMM field. The colored points show the eight ACTPol clusters with the redshift information. The gray points are HSC source galaxies used for our lensing analysis, which covers $29.5$~deg$^2$. Note that holes in the source galaxy distribution are due to the bright star mask.}
\label{fig:overlap}
\end{figure*}

Figure~\ref{fig:overlap} shows the overlap between the ACTPol D6 field and the HSC XMM field. There are eight clusters that span a redshift range of $0.186 \leq z \leq 1.004$. The average cluster redshift, which is weighted by the source galaxy weight described in Section~\ref{sec:weak_lensing_basics}, is $\langle z_l\rangle = 0.43$.

\begin{figure*}
  \begin{center}
    \includegraphics[width=\textwidth]{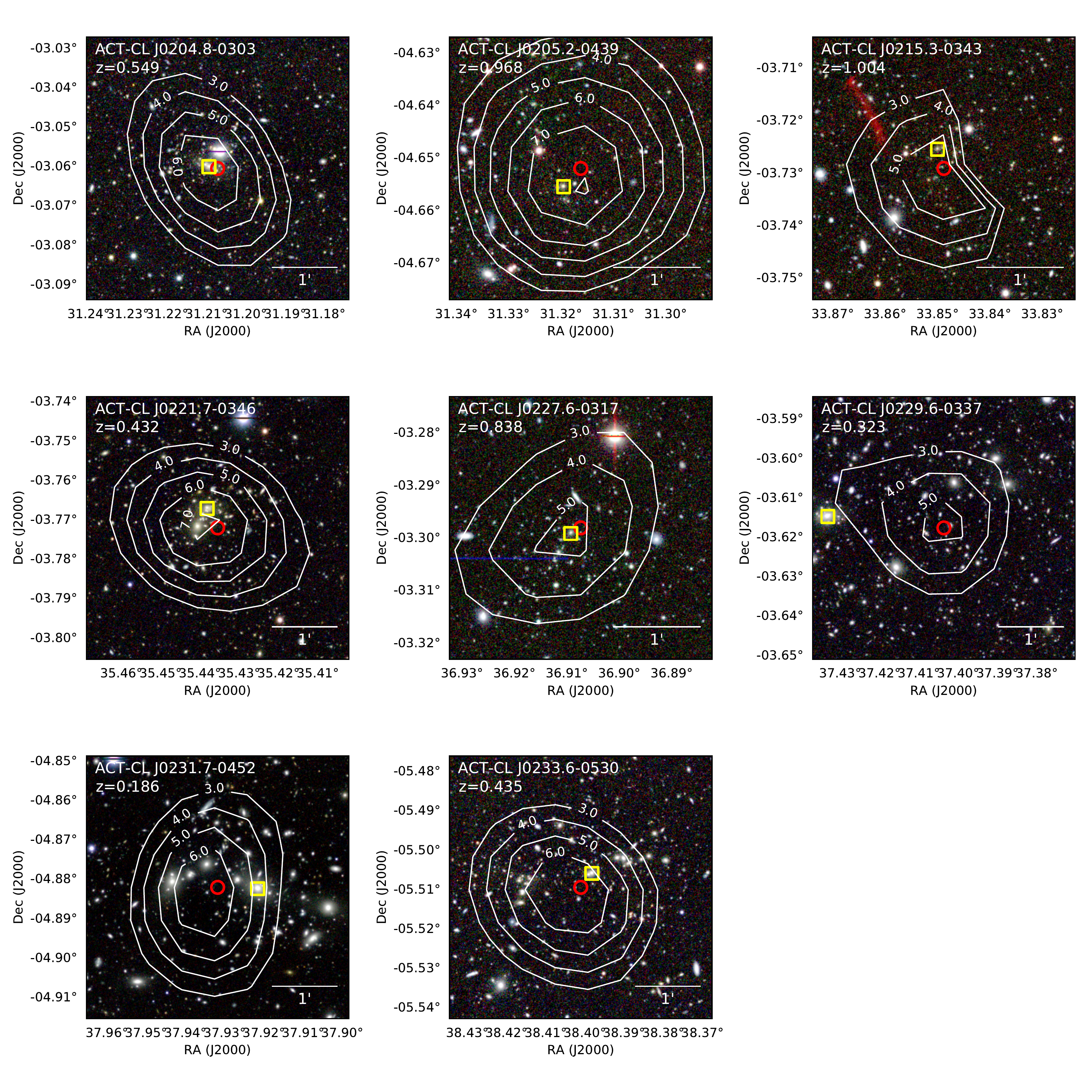}
  \end{center}
\caption{ACTPol SZ-selected clusters observed by HSC S16A wide XMM field. The images of clusters with $z>0.6$ are composed of $izy$-bands and the others are $riz$-bands. Red circles denote the SZ centers, and yellow squares denote the BCG centers. White contours show the SZ $S/N$ in units of $\sigma$
for the mm-wave detection. The distorted contours of ACT-CL J0215.3-0343 are due to a point-source mask.}
\label{fig:cluster_image}
\end{figure*}

Table~\ref{tab:cluster} lists the properties of the sample. When measuring the shear signal from the HSC data, we define the cluster centers as the brightest cluster galaxy (BCG) locations as determined by \citet{Hilton2017} using a combination of visual inspection and the $i$, $r-i$ color-magnitude diagram. We match the ACTPol clusters with optically-selected clusters in the HSC Wide S16A data set by the CAMIRA algorithm in \cite{Oguri:2018}, requiring separation $<  2^\prime$. The optical richness derived by CAMIRA, $N_\mathrm{cur}$, is shown in Table~\ref{tab:cluster}. Figure~\ref{fig:cluster_image} shows the HSC color images of our sample together with the SZ $S/N$ contours. The cluster centers which are defined by the BCG and SZ signals are consistent for most of the clusters, given that the beam of ACTPol is about $1.4 \arcmin$  at 148 GHz. Only ACT-CL J0229.6-0337 has a significantly large offset between these cluster centers. We will look into how the offset affects our lensing signal in Appendix \ref{off-centering}.

\begin{table*}
\caption{ACTPol clusters overlapping with the HSC XMM field.}
\begin{center}
\begin{tabular}{lcccccccc}
\hline
Name & SZ RA & SZ Dec & BCG RA & BCG Dec & Redshift$^1$ & SNR$_{\rm 2.4}$ & $M_{\rm SZ}$ [10$^{14}$M$_\odot$] & $N_{\rm cor}$$^2$\\
\hline
ACT-CL J0204.8-0303 & 2:04:49.73 & -3:03:38.42 & 2:04:50.27 & -3:03:36.82 & 0.549 & 6.84 & 3.04$^{+0.57}_{-0.48}$ & 35.0\\
ACT-CL J0205.2-0439 & 2:05:15.90 & -4:39:07.50 & 2:05:16.69 & -4:39:19.96 & 0.968 & 8.10 & 3.12$^{+0.52}_{-0.44}$ & 38.6\\
ACT-CL J0215.3-0343 & 2:15:23.72 & -3:43:45.20 & 2:15:24.01 & -3:43:31.98 & 1.004 & 5.86 & 2.46$^{+0.44}_{-0.37}$ & 44.8\\
ACT-CL J0221.7-0346 & 2:21:44.53 & -3:46:19.94 & 2:21:45.17 & -3:46:02.19 & 0.432 & 7.29 & 3.07$^{+0.59}_{-0.50}$ & 69.3\\
ACT-CL J0227.6-0317 & 2:27:37.77 & -3:17:53.48 & 2:27:38.22 & -3:17:57.31 & 0.838 & 5.15 & 2.19$^{+0.42}_{-0.35}$ & 50.9\\
ACT-CL J0229.6-0337 & 2:29:36.88 & -3:37:04.01 & 2:29:43.97 & -3:36:53.52 & 0.323 & 5.15 & 2.40$^{+0.54}_{-0.44}$ & 57.0\\
ACT-CL J0231.7-0452 & 2:31:43.63 & -4:52:56.16 & 2:31:41.17 & -4:52:57.44 & 0.186 & 6.85 & 3.08$^{+0.72}_{-0.58}$ & 116.4\\
ACT-CL J0233.6-0530 & 2:33:36.27 & -5:30:34.52 & 2:33:35.59 & -5:30:21.76 & 0.435 & 6.91 & 3.11$^{+0.61}_{-0.51}$ & 46.9\\
\multicolumn{5}{l}{$^1$ Redshifts are spectroscopic measurements.}\\
\multicolumn{5}{l}{$^2$ Richness is from the HSC CAMIRA cluster catalog \citep{Oguri:2018}.}
\end{tabular}
\end{center}
\label{tab:cluster}
\end{table*}

\subsection{HSC Source Galaxies}
\label{sec:HSC_source_galaxies}
HSC is the wide-field prime focus camera on the Subaru Telescope \citep{Miyazaki:2018,Komiyama:2018} located at the summit of Mauna Kea. The combination of the wide field-of-view (1.77~deg$^2$), superb image quality (seeing routinely less than $0.6^{\prime\prime}$), and large aperture of the primary mirror (8.2~m) makes HSC one of the best instruments for conducting weak-lensing cosmology. Under the Subaru Strategic Survey Program \citep[SSP;][]{Aihara:2018a}, HSC started a galaxy imaging survey in 2014 that aims to cover 1,400~deg$^2$ of the sky down to $i = 26$ after its 5 years of operation. The first-year galaxy shape catalog \citep{Mandelbaum:2018} was produced using the data taken from March 2014 through April 2016 with about 90 nights in total. The first-year data consists of six distinct fields (HECTOMAP, GAMA09H, WIDE12H, GAMA15H, XMM, VVDS) and covers $136.9$~deg$^2$ in total. Note that this catalog is a slight extension of Data Release 1 \citep{Aihara:2018b}. As mentioned above, we use the shape catalog in the XMM field ($29.5$~deg$^2$ once we remove the star mask region) which overlaps with the current ACTPol observations. The weighted number density of source galaxies in this field is 22.1~arcmin$^{-2}$ and their median redshift is $z_{\rm m} =0.82$.

Here we briefly summarize the HSC shape catalog. For details of the shape catalog production, see \cite{Mandelbaum:2018}. The galaxy shapes were estimated on coadded $i$-band images by the re-Gaussianization technique \citep{Hirata2003}, which is a moment-based method with PSF correction. This method was extensively used and characterized in the Sloan Digital Sky Survey \citep{Mandelbaum2005, Reyes2012, Mandelbaum2013}. The shapes $(e_1, e_2)=(e\cos 2\phi, e\sin 2\phi)$, where $\phi$ is position angle, are defined in terms of {\it distortion}, $e=(a^2-b^2)/(a^2+b^2)$, where $a$ and $b$ are the major and minor axes, respectively \citep{BJ2002}. The galaxy shapes were calibrated against image simulations generated with \textsc{GalSim} \citep{Rowe2015}, an open source software package, which yields correction factors for the shear measurements. These factors are the multiplicative bias $m$ and additive bias $(c_1, c_2)$, which are defined as $g_{i, \rm{obs}}=(1+m)g_{i, \rm{true}} + c_i$, where $(g_1, g_2)$ is defined in terms of {\it shear}, $g=(a-b)/(a+b)$, and must be applied to the shear measurements. Note that the multiplicative bias is shared between the two ellipticity components \citep[for details, see][]{Mandelbaum:2017}. For each galaxy, the shape catalog provides an estimate of the intrinsic shape noise, $e_{\rm rms}$, an estimate of measurement noise, $\sigma_e$, and inverse weights $w$ from combining $e_{\rm rms}$ and $\sigma_e$. The measurement noise is statistically estimated from the shape measurements performed on simulated images, and the intrinsic shapes are derived by subtracting the measurement noise from the ellipticity dispersion measured from the real data. Note that we use a catalog made with the ``Sirius" star mask, which actually includes bright galaxies and thus an extended region around each BCGs may be masked as well. After we performed the lensing measurement, the shape catalog was updated with a more reliable star mask called ``Arcturus" \citep[for details, see][]{Coupon:2018}. We checked that switching to this new shape catalog changes our fiducial stacked weak-lensing measurements by an amount well within the statistical uncertainty (typically 10~\%).

We use photometric estimates to select source galaxies based on colors. For this purpose, we use \textsc{cmodel} magnitudes derived by fitting a galaxy's light profile with a composite model of the exponential and de Vaucouleurs profile \citep{Bosch:2018}. The HSC SSP catalog has photo-$z$ estimates based on six different methods \citep{Tanaka:2018}. Among these methods, we use MLZ, an unsupervised machine learning method based on the Self-Organizing Map which is a projection map from multi-dimensional color space to redshift, for our fiducial measurement. We have checked the consistency among lensing signals with different photo-$z$ methods, which is described in Appendix~\ref{app:different_photozs}. In this paper, we use the redshift PDFs, $P(z)$, and randomly sampled point estimates that are drawn from the PDFs, $z_{\rm mc}$. The latter is specifically used for one of the source galaxy selection methods described below.

\section{Weak-Lensing Measurements}
\label{sec:weak_lensing}
In this section, we describe basics of weak lensing measurement in Section~\ref{sec:weak_lensing_basics}, covariance estimation in Section~\ref{sec:covariance}, and key systematic tests, i.e., source galaxy selection and photo-$z$ bias in Sections~\ref{sec:source_galaxy_selection} and \ref{sec:photo-z_bias}. Additional systematic tests we performed are described in Appendix~\ref{app:systematc_tests}.

\subsection{Weak Lensing Basics}
\label{sec:weak_lensing_basics}
Weak gravitational lensing manifests as a coherent distortion of apparent shapes of source galaxies. For a source galaxy at a comoving transverse separation $R$ from the lens center, the lens' gravitational potential induces a tangential distortion, $\gamma_t$, which depends on the lens' matter density profile projected along the line of sight, $\Sigma(R)$, and on the redshifts of the source galaxy, $z_s$, and the lens, $z_l$. For the purposes of this work the lenses are galaxy clusters. In terms of the average projected mass density inside $R$, $\bar{\Sigma}(< R)$, and the critical surface mass density $\Sigma_{cr}(z_l, z_s)$ defined below, the tangential distortion can be expressed in terms of the excess surface mass density $\Delta\Sigma(R)$ as follows:

\begin{equation}
\gamma_t(R)=\frac{\bar{\Sigma}(<R)-\Sigma(R)}{\Sigma_{\rm cr}(z_l, z_s)}\equiv\frac{\Delta\Sigma(R)}{\Sigma_{\rm cr}(z_l, z_s)}.
\end{equation}
In terms of the angular diameter distances of the source and lens from us, $D_A(z_s)$ and $D_A(z_l)$, and the angular diameter distance between the two, $D_A(z_s, z_l)$, $\Sigma_{\rm {cr}}(z_s,z_l)$ is defined as: 
\begin{equation}
\Sigma_{\mathrm{cr}}(z_l,z_s) = \frac{c^2}{4\pi G} \frac{D_A(z_s)}{(1+z_l)^2D_A(z_l)D_A(z_l,z_s)}.
\end{equation}
where the factor $(1+z_l)^{-2}$ comes from our use of comoving coordinates \citep{Mandelbaum2006}.

To estimate cluster properties including mass from measurements of $\Delta\Sigma$, we will start with models of the cluster density radial profile, $\rho(r)$, and integrate to generate modeled lensing profiles $\Delta\Sigma(r; M,x)$, where $x$ signifies other possible parameters of the models (see Section~\ref{sec:results}). From the data, we can estimate $\Delta\Sigma$ in each radial bin $R_i$  for either a single cluster or a stack of multiple clusters with a weighted sum over the tangential components of  the shapes of galaxies, $e_t$, as follows. We use a weighting based both on the shape catalog weight for the source galaxy, $w_s$, and on an estimate of the appropriate critical surface mass density for the particular source-lens pair, using the photo-$z$ PDF for each source, $P_s(z)$, to account for the dilution effect of foreground galaxies. (See Section~\ref{sec:source_galaxy_selection} for discussion of the impact of possible cluster member contamination, however.)  Specifically, we define $\tilde{w}_{ls} \equiv w_s  \langle \Sigma_{cr,ls}^{-1}\rangle^2$, where
\begin{equation}
\label{eq:critical_surface_mass_density}
\left\langle \Sigma_{{\rm cr}, ls}^{-1} \right\rangle = \frac{\int_0^\infty P_s(z_s) \Sigma_{{\rm cr}}^{-1}(z_l, z) dz}{\int_0^\infty P_s(z) dz}.
\end{equation}
With two additional calibration factors described below, the lensing profile is estimated as:
\begin{equation}
\Delta\Sigma(R_i) = \frac{1}{{2\cal R}(R_i)}\frac{\sum_{ls \in R_i} \tilde{w}_{ls} e_{t,ls} \left\langle \Sigma_{{\rm cr}, ls}^{-1} \right\rangle^{-1}}{(1+K(R_i))\sum_{ls \in R_i} \tilde{w}_{ls}}.
\label{eq:stack}
\end{equation}

The factor $1+K(R)$ is the shear calibration factor which corrects for multiplicative bias described in Section~\ref{sec:HSC_source_galaxies};
\begin{equation}
1+ K(R_i) = \frac{\sum_{ls \in R_i} \tilde{w}_{ls} \left(1+ m_{s}\right)}{\sum_{ls \in R_i} \tilde{w}_{ls}},
\end{equation}
where $m_{s}$ is the multiplicative bias of the source, $s$. The shear responsivity ${\cal R}(R_i)$ is necessary to take into account the summation in non-Euclidean shear space. It is calculated as
\begin{equation}
{\cal R}(R_i) = 1-\frac{\sum_{ls \in R_i }\tilde{w}_{ls} e_{{\rm rms}, ls}^2}{\sum_{ls \in R_i} \tilde{w}_{ls}}.
\end{equation}
We compute $\Delta\Sigma$ in 12 logarithmic bins from 0.1~$\mpch$ to 10~$\mpch$. Note that we do not use all the radial bins for model fitting, as described later.

\begin{figure*}
  \begin{center}
    \includegraphics[width=\textwidth]{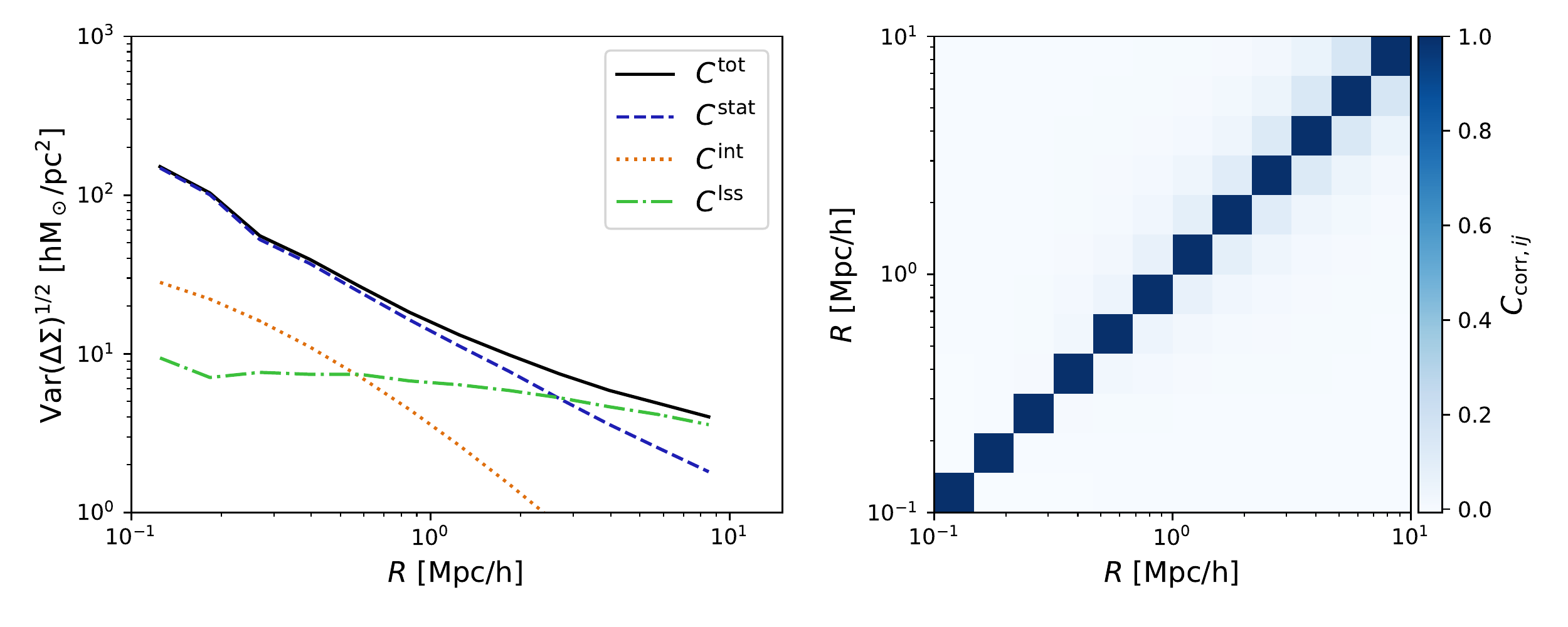}
  \end{center}
\caption{{\it Left:} Diagonal components of the covariance matrix used in the stacked analysis. The black solid curve denotes the total covariance; the blue dashed curve denotes the statistical uncertainty due to intrinsic shapes of source galaxies, the orange dotted line denotes the covariance due to intrinsic variations of cluster properties such as triaxiality and concentration, and the green dashed-dotted line denotes the covariance due to large scale structure uncorrelated with clusters.
Shape noise is dominant for $r < 2.8 \mpch$, while the large-scale structure covariance dominates at larger separations.
{\it Right:} The stacked analysis correlation matrix of the total covariance as a function the radial bin. The correlation between radial bins appears at large scales due to the large scale structure covariance.}
\label{fig:covariance}
\end{figure*}
\subsection{Covariance}
\label{sec:covariance}
We estimate the covariance matrix of the lensing signal as
\begin{equation}
C = C^{\rm stat} + C^{\rm int} + C^{\rm lss},
\end{equation}
where $C^{\rm stat}$ is the statistical uncertainty due to galaxy shapes:
\begin{equation}
C^{\rm stat}_{ij}=\frac{1}{4{\cal R}^2(R_i)}\frac{\sum_{ls \in R_i} \tilde{w}^2_{ls} (e^2_{{\rm rms},ls} + \sigma_{e, ls}^2)\left\langle \Sigma_{{\rm cr}, ls}^{-1} \right\rangle^{-2}}{\left[1+K(R_i)\right]^2\left[ \sum_{ls \in R_i} \tilde{w}_{ls}\right]^2}\delta_{ij};
\end{equation}
$C^{\rm int}$ accounts for the intrinsic variations of projected cluster mass profiles such as halo triaxiality, the presence of correlated halos, and the intrinsic scatter of the concentration--mass relation \citep{Gruen2015}; and $C^{\rm lss}$ is due to uncorrelated large scale structure (LSS) along the line of sight. Detailed calculations of the intrinsic and LSS covariances are described in Appendix~\ref{app:covariance}. Figure~\ref{fig:covariance} shows the diagonals of the covariance matrix used in our analysis and the correlation matrix, defined as $C_{{\rm corr},ij} = C_{ij}/\sqrt{C_{ii}C_{jj}}$. A similar figure was presented in \citet{Umetsu2016}, which describes a joint weak and strong lensing analysis of 20 high-mass clusters. Figure~\ref{fig:covariance} shows that the total uncertainty is dominated by the shape noise ($C^{\rm stat}$) at $r\simlt 3$\,Mpc\,$h^{-1}$, beyond which the relative contribution from 
LSS noise, uncorrelated with the cluster, becomes important.

\subsection{Source Galaxy Selection}
\label{sec:source_galaxy_selection}
If cluster galaxies are misidentified as background galaxies, they will introduce a systematic dilution of the weak-lensing signal from their galaxy cluster. We look into two distinct source galaxy selection methods which were established in \cite{Medezinski:2017} with the CAMIRA catalog of optically-selected clusters in HSC SSP: a selection based on the color-color space (the CC cut) and another based on a cumulative photo-$z$ PDF (the $P(z)$ cut). Note that we use MLZ to define the latter cut and calculate lensing signals.

The CC cut is defined in the $g-i$ vs $r-z$ space to minimize the dilution in lensing signal due to the contamination by cluster members and foreground galaxies. The CC cut is defined differently for a cluster with $z_l\leq0.4$ and $z_l>0.4$ to avoid excessively removing galaxies behind low redshift clusters (for the detailed definition, see Appendix A in \citealt{Medezinski:2017}). Using the CC cut \citealt{Medezinski:2017} showed that the dilution is consistent with zero.

The $P(z)$ cut initially proposed in \citet{Oguri2014} is defined by two criteria that each galaxy must satisfy to be identified as a background galaxy. The first criterion is
\begin{equation}
p_{\rm cut} < \int_{z_{\rm min}}^\infty P(z){\rm d}z,
\end{equation}
where $p_{\rm cut} = 0.98$, meaning that we require that 98\% of the area beneath the $P(z)$ lies beyond $z_{\rm min}$. For our analysis $z_{\rm min} = z_l+ \Delta z$ and we employ $\Delta z=0.2$ for a secure rejection of cluster galaxies, following the investigation by \cite{Medezinski:2017}. The second criterion is that each galaxy's randomly drawn point redshift value from its photo-z PDF, $z_{\rm mc}$, be less than $z_{\rm max}$. This criterion rejects photo-z PDFs that are predominantly above the redshift limit that are considered secure for a given optical survey. This maximum redshift is optimized for HSC and set to $z_{\rm max}=2.5$ \citep[see][]{Medezinski:2017}.

\begin{figure}
\includegraphics[width=\columnwidth]{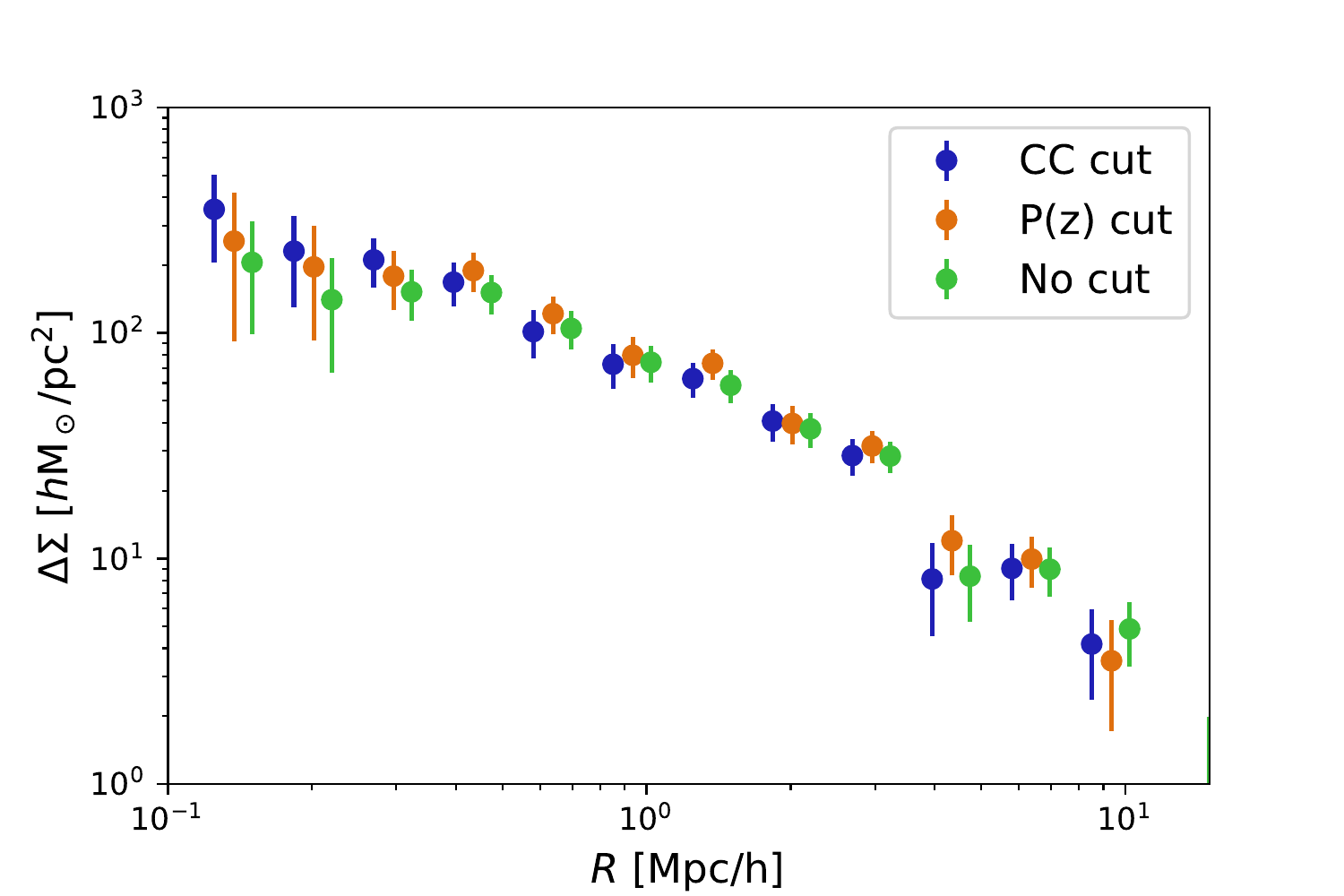}
\caption{The stacked weak-lensing signal from the eight clusters in the sample for different source galaxy selection methods to test the contamination from galaxies physically associated with clusters. The data points with the $P(z)$ cuts and no cuts are shifted along the x-axis for illustrative purposes. The CC cut denotes the selection in the color-color space detailed in \cite{Medezinski:2017}, while the $P(z)$ cut is based on \cite{Oguri2014} with some tweaks. We also show the signal without any source selection. The dilution effects due to contamination by foreground galaxies are already corrected for because of our use of full photo-$z$ PDF in the calculation of critical surface mass density (see Eq.~\ref{eq:critical_surface_mass_density}). We use the CC cut for our fiducial measurement.}
\label{fig:lensing_source_selection}
\end{figure}

Figure~\ref{fig:lensing_source_selection} compares the stacked lensing signal from the eight clusters in the sample calculated without source selection cuts, with CC cuts, and with P(z) cuts. We do not find a significant difference among these signals within the error bars. Following the extensive analysis in \cite{Medezinski:2017} that found the CC cuts to be less diluted, we use the CC cut for our fiducial measurement.

\subsection{Photo-$z$ Biases}
\label{sec:photo-z_bias}
Systematic biases in the photometric redshift estimates of source galaxies would propagate to the weak lensing signal measurement through the calculation of the critical surface density. Following \cite{Mandelbaum2008}, this bias in the weak lensing signal of a cluster at redshift $z_l$ can be estimated as 
\begin{equation}
\frac{\Delta\Sigma}{\Delta\Sigma^{\rm true}} (z_l) = 1 + b(z_l) = \frac{\sum_s \tilde w_{ls} \langle \Sigma_{{\rm cr},ls}^{-1} \rangle^{-1} [\Sigma_{{\rm cr},ls}^{\rm true}]^{-1}}{\sum_s \tilde w_{ls}}\,,
\label{eq:photoz_bias}
\end{equation}
where the quantities with a superscript "true" denote the quantity as would be measured with a spectroscopic sample, the sum over $s$ goes over all source galaxies.

Nominally such photo-z biases are evaluated using a spectroscopic redshift (spec-$z$) sample that is independent from those used to calibrate the photometric redshifts and has the same population properties (magnitude and color distribution) as our source redshift sample. In practice it is difficult to obtain such a representative spec-$z$ sample given the depth of our source catalog. In principle the difference among the populations of an existing spec-$z$ sample and the weak lensing source sample can be accounted for by using a clustering and reweighting technique \citep[for assumptions and caveats of this method, see][]{Bonnett2016,Gruen2017}. This method decomposes galaxies in the source sample into groups with similar properties. Then the galaxies from the spectroscopic sample in these groups are reweighted to mimic the distribution of the weak lensing source sample.

The first-year HSC shape catalog has substantial overlaps with public spec-$z$ samples, such as GAMA and VVDS, however, they are not enough galaxies with spec-z to represent the source sample even after reweighting. Instead of the spec-$z$ sample we use the COSMOS-30 band photo-$z$ \citep{Ilbert2009} sample. We decompose the galaxies in the weak lensing sample using their $i$ band magnitude and 4 colors into cells of a self-organizing map \citep[SOM,][]{Moreprep}. Using the HSC photometry of the COSMOS 30-band photo-$z$ sample, we classify them into SOM cells defined by the source galaxy sample and compute their new weights ($w_{\rm SOM}$) which adjusts the COSMOS 30-band photo-$z$ sample to mimic our source galaxy sample. We compute the photo-$z$ bias (see Eq.~\ref{eq:photoz_bias}) by including $w_{\rm SOM}$ in the definition of $w_{ls}$. Then we average Eq.~\ref{eq:photoz_bias} over our cluster sample with the weight defined by Eq.~23 in \cite{Nakajima2012}.  This yields a two percent bias which negligible compared to statistical uncertainties.

\section{Results}
\label{sec:results}
\subsection{Individual Cluster Measurements}
\label{sec:indv}
\begin{figure*}
\includegraphics[width=\textwidth]{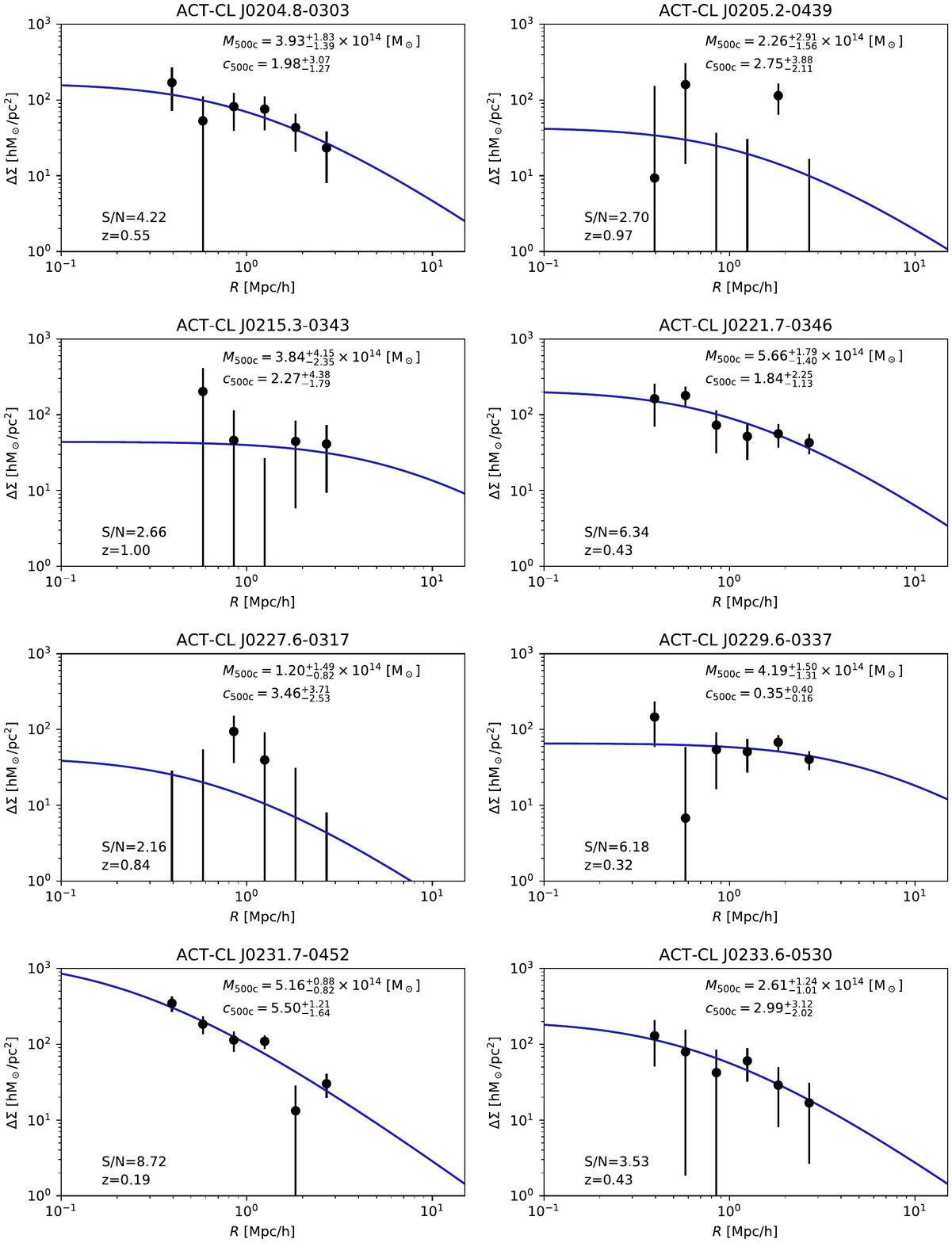}
\caption{The weak-lensing signals for individual clusters. The blue solid curves denote the best-fit NFW profile for which we use data points in the radial range of $0.3 \mpch < R < 3 \mpch$. The median and 68\% confidence levels of $M_{\rm 500c}$ and $c_{\rm 500c}$ are show in the upper right of each panel. The signal-to-noise ratio of the lensing measurement and the cluster redshift are shown in the lower left.}
\label{fig:individual_lensing}
\end{figure*}

In Figure~\ref{fig:individual_lensing}, we show the lensing signal, $\Delta \Sigma$, for each cluster in our sample.  We estimate properties for each cluster using the Navarro-Frenk-White (NFW) radial density profile \citep{NFW1996,NFW1997}: 
\begin{equation}
\rho_{{}_{\rm NFW}}(r)=\frac{\rho_s}{(r/r_s)(1+r/r_s)^2}.
\end{equation}
Following \citet{Okabe2010}, we convert the parameterization of $\rho_{{}_{\rm NFW}}$ from characteristic density and  scale, $\rho_s$ and $r_s$,  to mass $M_{200m}$, and concentration, defined as $c_{200m} \equiv R_{200m}/r_s$.   We integrate the 3D $\rho_{{}_{\rm NFW}}$  to generate the fitting function $\Delta\Sigma($R$; M_{200m}, c_{200m})$ and fit for both parameters. We restrict the fitting range to $0.3\mpch < R < 3\mpch$, where the lower limit is to avoid using blended images as blending is more prominent towards the cluster center \citep{Medezinski:2017}, and the upper limit is to avoid the fitting being affected by the 2-halo regime \citep{Applegate2014}. We use only shape noise for this fitting, since the other components are negligible at these scales (see Figure~\ref{fig:covariance}). The constraint on the enclosed mass and concentration are then converted to $M_{\rm 500c}$ and $c_{\rm 500c}$, as shown in each panel of Figure~\ref{fig:individual_lensing} along with the signal-to-noise ratio for the weak-lensing measurement and the cluster redshift. The three high-redshift clusters, ACT-CL J0205.2-0439, ACT-CL J0215.3-0343, and ACT-CL J0227.6-0317, have low signal-to-noise ratios which are reflected in their poor best fit curves and their weakly constrained lensing mass and concentration. 

\subsection{Stacked Cluster Measurement}
\label{sec:stacked_wl}
We obtain the stacked cluster lensing signal (Equation~\ref{eq:stack}) with a signal-to-noise ratio of $11.1$ ($9.6$ for the data used for mass inference, see below). Our goal is to estimate the average mass of the clusters in the stack to compare to the average $M_{SZ}$. We introduce an improved method for estimating the average mass in Section~\ref{section:complete_stacked_model}, focusing on two models for the density profiles, as described below.  For comparison with earlier work, in Section~\ref{sec:single_mass-bin_fit} we also estimate the average mass with the single-mass-bin fit, which is done using only the stacked cluster signal, fitted with a model parameterized by a single mass.

The single-mass-bin fit has been widely performed in the literature, but is limited by the fact that the amplitude of the lensing signal is not linearly proportional to the cluster mass.  Our method improves upon it by emulating the stacking process and incorporating the effects of the cluster mass function and selection function.

For estimating the average cluster mass, we use the ``Dark Emulator" model and the ``Baryonic Simulations" model described in the next few paragraphs.  We also present results from the NFW density profile for comparison to earlier work. Both these models provide excess surface density profiles up to the two-halo regime, allowing use of the large scale information to constrain halo mass. Thus, we extend the fitting range up to  $10 h^{-1}$ Mpc for those models, rather than the $3 h^{-1}$~Mpc limit for the NFW fit, which results in tighter constraints on cluster mass.

The Dark Emulator model is based on a cosmic emulator developed by \cite{Nishimichi2017inprep} and it predicts statistical quantities of halos, including the mass function, the halo-matter cross-correlation, and the halo autocorrelation, as a function of halo mass, redshift, and cosmological model.  The Dark Emulator model is based on a large set of $N$-body simulations; details can be found in \cite{Murata2017}.  We convert $M_{200m}$ to $M_{500c}$  assuming the NFW profile with $c_{200m}$ derived by \cite{Diemer2015}. To compute the concentration and conversion, we use the public, open source software called \textsc{Colossus} \citep{Diemer2017}.

The Baryonic Simulations model is based on hydrodynamical simulations of cosmological volumes described in \citet{BBPSS}. From these simulations we project the mass distributions of the halos in all the simulation snapshots at given redshifts following the methodology in \citet{Battaglia2016}. 

\subsubsection{Single Mass-bin Fit}
\label{sec:single_mass-bin_fit}

\begin{figure*}
\begin{center}
\includegraphics[width=\columnwidth]{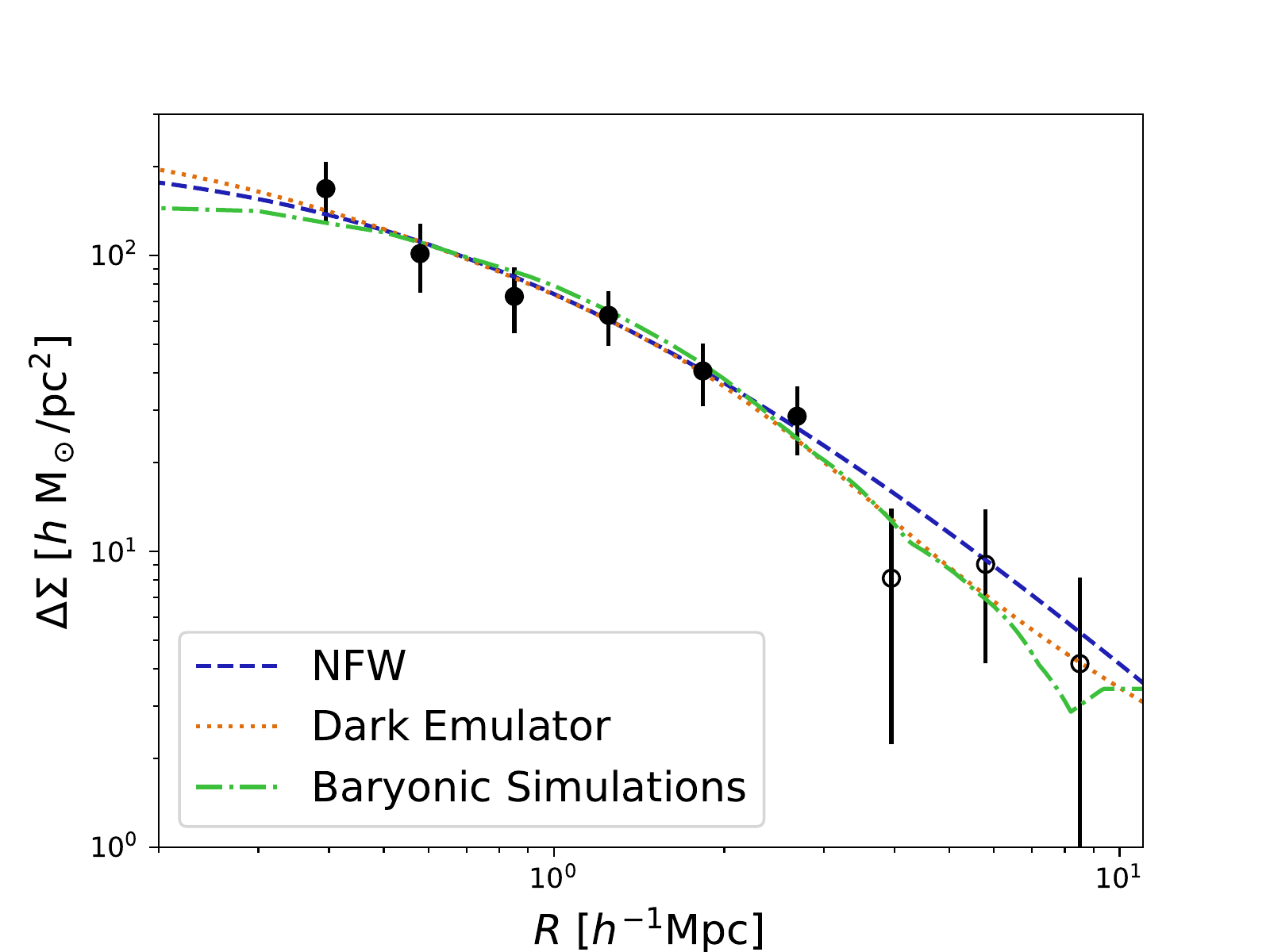}
\includegraphics[width=\columnwidth]
{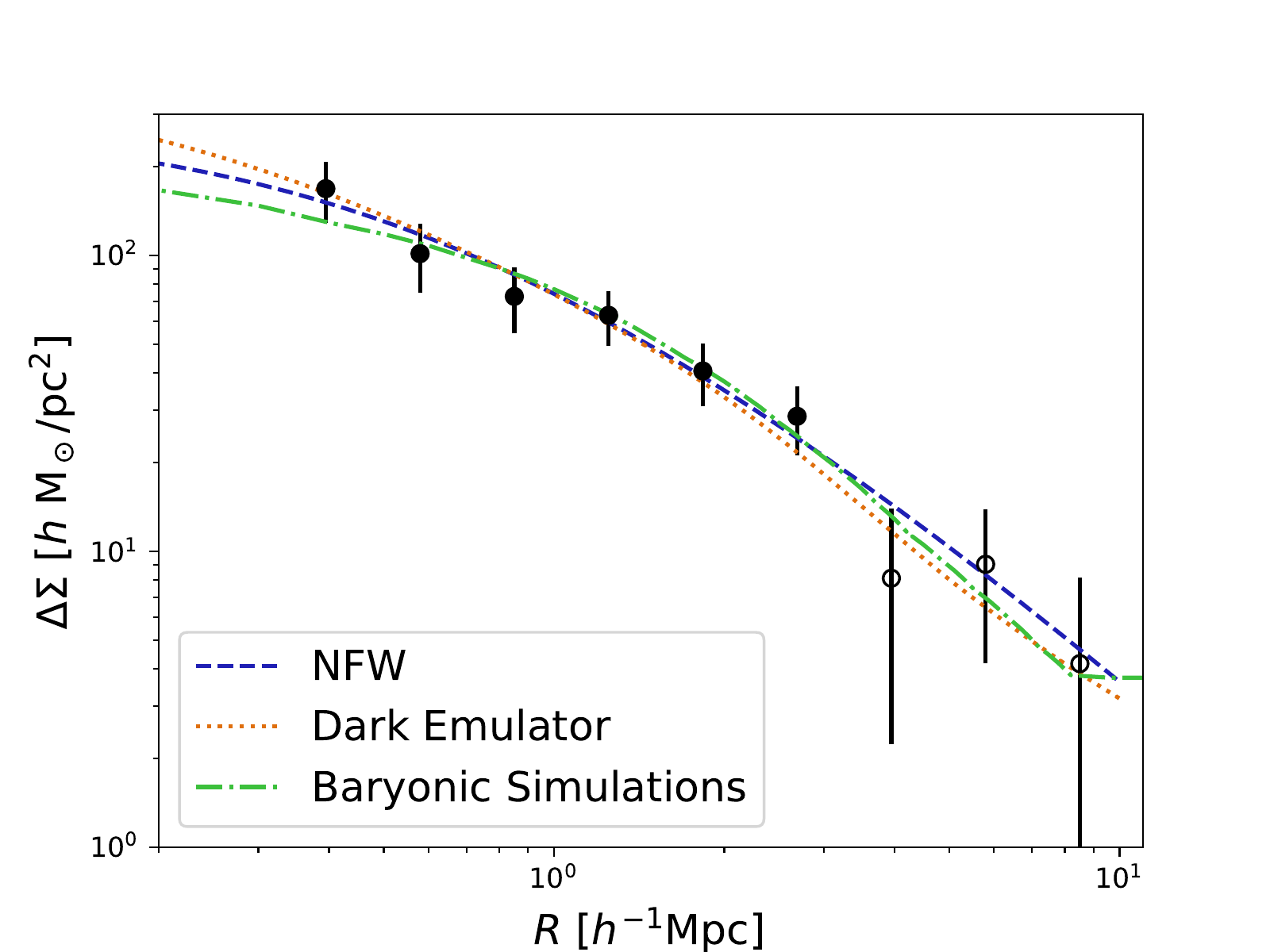}
\end{center}
\caption{{\it Left:} Single mass-bin fit on the stacked lensing measurement. {\it Right:} Stacked model fit on the stacked lensing measurement. The filled circles show data points used for the NFW fit, while both the filled and open circles are used for the Dark Emulator and Baryonic Simulations models.}
\label{fig:stacked_lensing}
\end{figure*}

The results of the single-mass-bin fits are summarized in Table~\ref{tab:stacked_lensing_stacked_model_fit} and the model fits compared to measurements are shown in Figure~\ref{fig:stacked_lensing}. When fitting the NFW profile, we again vary both $M_{200m}$ and $c_{200m}$. We assume all the clusters are at a single redshift, which we calculate as a weighted average over the lens-source pairs used in the stacked measurement:  $ \langle z_l \rangle_{{\rm stack}} = \sum_{ls} \tilde{w}_{ls} z_l / \sum_{ls} \tilde{w}_{ls} = 0.43.$ For the Baryonic Simulations model, we average the calculated excess surface density profile as a function of mass in the three redshift outputs that are closest to the value $ \langle z_l \rangle_{{\rm stack}}$. As Table~\ref{tab:stacked_lensing_stacked_model_fit} shows, the single-mass-bin fits have reasonable $\chi^2$ values (here we fixed cosmological and other models parameters, thus reducing the
number of free parameters to just two for the NFW fit, mass and concentration) and yield cluster masses that are within the $1\sigma$ errors. However, these models are being applied to the same data, so the differences among the inferred masses cannot be due to statistical error. We interpret these differences to be from systematic errors resulting from modeling uncertainty. Note that since the Dark Emulator is a more complete DM-only model for the mass profile than the NFW profile, we will focus on comparing it to the Baryonic Simulations model. We will continue to show the NFW fit results so that our results can be compared with previous results that did not use emulator fits. We find that using the single-mass-bin fit introduces systematic modeling uncertainty of  order 15\% for this sample of clusters, which is comparable to the statistical uncertainty for this cluster sample.

\subsubsection{Stacked Model Method}
\label{section:complete_stacked_model}
Here we describe our improved method for estimating the average cluster mass $\langle M_\mathrm{WL} \rangle$ from the cluster sample, which we call the Stacked Model method. The essence of the method is to model the stacked lensing profile $\langle \Delta \Sigma \rangle (R)$, accounting for:
\begin{itemize}
\item the weak-lensing weighting, which depends on the cluster $z$; \item the cluster (halo) mass function, $dn(z)/dM$;
\item the ACTPol cluster selection function, which is known in terms of $M_{SZ}$ and $z$;  and
\item the mass-mass scaling relationship between $M_{SZ}$ and $M$.
\end{itemize}
In particular, we define the variables $\mu_{\rm SZ} \equiv {\rm ln} M_{\rm SZ}$ and $\mu \equiv {\rm ln} M$, and we assume
\begin{equation}
\mu_{\rm SZ} = \ B\, \mu + A,
\end{equation}
but fix the mass-dependent exponent $B$ to unity, so that we can focus on $A$, which quantifies the constant mass bias between $M_{\rm SZ}$ and the true mass $M$. We qualitatively discuss mass dependence in Section~\ref{sec:massbias}, but the $S/N$ of our measurement prevents us from providing an interesting constraint on $B$. We consider cluster density profiles from the models described in Section~\ref{sec:stacked_wl}. To construct a model that can be fit to the data as a function of stacked cluster mass, we compute the stacked lensing profile and stacked mass for different values of $A$, and then interpolate the stacked lensing profile as a function of stacked mass.

The weak-lensing weights are computed for the redshifts $z_j$ of the eight clusters as $w_l(z_j,R_i) = \sum_s \tilde{w}_{ls}(R_i)$,  where $s$ includes the subset of source galaxies for the $j$th cluster after the CC cut (Section 3.2).  The weak-lensing weights are a smooth function of $z$, and so $w_l(z,R_i)$ is estimated by extrapolation from $w_l(z_j,R_i)$ for the Dark Emulator and Baryonic Simulations models. The inherent $dn(z)/dM$ from the simulations are self-consistently used for the Dark Emulator and Baryonic Simulations models and for the NFW profile modeling the $dn(z)/dM$ from the Dark Emulator is used. The cluster selection function, $S(M_{\rm SZ}, z)$, is computed by averaging the ACTPol survey completeness map, which is defined for a given SZ mass and redshift under our selection criteria SNR$_{2.4} > 5$, over the XMM field where the HSC source galaxies exist (see Section~\ref{sec:HSC_source_galaxies} for details). 

\begin{table*}
\caption{Parameter constraints from the Single Mass-bin fit and Stacked Model method fit to the stacked lensing data.}
\begin{center}
\begin{tabular}{llccc}
\hline
&Parameter & NFW & Dark Emulator & Baryonic Simulation\\
\hline
Single Mass-bin & & & & \\
\hline
&$M_{\rm WL}$ [10$^{14}$M$_{\odot}$] & $4.26^{+0.82}_{-0.71}$ & $4.22^{+0.72}_{-0.64}$ & $3.67^{+0.86}_{-0.58}$ \\
&$c_{\rm 500c}$ & $2.08^{+0.86}_{-0.71}$ & N/A & N/A\\
&$\chi^2$/dof & 1.3/4 & 2.5/8  & 3.4/8 \\
\hline
Stacked Model & & & & \\
\hline
&$\langle M_{\rm WL} \rangle$ [10$^{14}$M$_{\odot}$] & $4.02^{+0.65}_{-0.61}$& $3.89^{+ 0.61}_{-0.57}$ & $3.55^{+0.63}_{- 0.48}$\\
&$1-b$ & $0.71^{+0.13}_{-0.12}$ & $0.74^{+0.13}_{-0.12}$ & $0.80^{+ 0.16}_{-0.12}$\\
&$\chi^2$/dof   & $1.6/5$ & $3.0/8$ & $3.1/8$\\
\hline
\end{tabular}
\end{center}
\label{tab:stacked_lensing_stacked_model_fit}
\end{table*}

We estimate the conditional probability distribution for $M_{SZ}$ given $M$ assuming it follows a log-normal distribution, such that
\begin{equation}
P(\mu_{\rm SZ}|\mu) = \frac{1}{\sqrt{2\pi}\sigma_{\mu_{\rm SZ}|{\mu}}}\exp{\left[-\frac{\left(\mu_{\rm SZ} - \mu - A\right)^2}{2\sigma^2_{{\mu_{\rm SZ}|\mu}}}\right]},
\end{equation}
where $\sigma_{\mu_{\rm SZ}|{\mu}}$ is the SZ mass-mass scatter which we fix to $\sigma_{\mu_{\rm SZ}|{\mu}}$ = 0.2 for the following reasons.  First, the lensing signal does not constrain the scatter well \citep[e.g.,][]{Murata2017}. Second, this is the same scatter assumed to correct the Eddington bias for the $M_{\rm SZ}$ values quoted in \citet{Hilton2017} and was used in \citet{Battaglia2016}. Thus, using this value for $\sigma_{\mu_{\rm SZ}|{\mu}}$ allows for direct comparison to previous results from ACT.

The Stacked Model lensing signal used for fitting is:
\begin{eqnarray}
\langle \Delta\Sigma \rangle (R_i) &=& \frac{1}{n_{\rm SZ}}\int dz w_l(z) \frac{cr^2(z)}{H(z)}\int d\mu M \frac{dn}{dM}(z) \nonumber \\
&&\int  d\mu_{\rm SZ} M_{\rm SZ} S(M_{\rm SZ}, z) P(\mu_{\rm SZ}|\mu)\Delta\Sigma(R_i,M,z),\nonumber \\
\label{eq:delsigstack}
\end{eqnarray}
Here $r^2(z)c/H(z)$ is the comoving volume per unit redshift interval and per unit steradian, $\Delta\Sigma(M,z, R_i)$ depends on the cluster profile model considered, and $n_{\rm SZ}$  is the expected number density per unit steradian of ACTPol clusters given the weak-lensing weights:
\begin{eqnarray}
n_{\rm SZ} &=& \int dz w_l(z) \frac{cr^2(z)}{H(z)} \int d\mu M \frac{dn}{dM}(z) \nonumber\\ 
&&\int d\mu_{\rm SZ} M_{\rm SZ} S(M_{\rm SZ}, z)P(\mu_{\rm SZ}|\mu).
\end{eqnarray}
Finally, we estimate the average mass for the cluster sample as:
\begin{eqnarray}
\langle M_{\rm WL} \rangle &=& \frac{1}{n_{\rm SZ}}\int dz w_l(z) \frac{cr^2(z)}{H(z)}\int d\mu M \frac{dn}{dM}(z) \nonumber \\
&&\int  d\mu_{\rm SZ} M_{\rm SZ} S(M_{\rm SZ}, z) P(\mu_{\rm SZ}|\mu)M_{\rm WL}(M,z).\nonumber\\
\label{eq:mfivestack}
\end{eqnarray}

For the different models of $\Delta\Sigma(R_i,M,z)$ there are subtle implementation differences that are specific to that model. For the NFW profile, we use the concentration-mass relation derived by \cite{Diemer2015}. Since the mass function in Dark Emulator is defined as a function of $M_{\rm 200m}$, we convert from $M_{\rm 500c}$ to $M_{\rm 200m}$ in the selection function with the concentration-mass relation defined above. For the Dark Emulator profile, we compute the conversion as described in Section \ref{sec:single_mass-bin_fit}.

For the Baryonic Simulations model we calculate the stacked weak-lensing signal as calculated in \citet{Battaglia2016} for a given average sample mass (Eqs.~\ref{eq:delsigstack} and~\ref{eq:mfivestack}) for each simulated halo surface density profile by the weak-lensing weight, the volume factor (comoving distance squared) associated with each simulation snapshot, the scatter in the scaling relation, and the ACTPol selection function to the simulated halos, described in \citet{Hilton2017}.

We summarize the results of fits for $M_\mathrm{WL}$ in Table~\ref{tab:stacked_lensing_stacked_model_fit} and show the best fit profiles in Figure~\ref{fig:stacked_lensing}. Similar to the single-mass-bin fit, we restrict the fitting range of the NFW fit to $0.3\mpch < R < 3\mpch$, but using the larger radii up to $R\sim10\mpch$ changes the constraint well within the $1\sigma$ statistical uncertainty. The resulting masses from the Dark Emulator and Baryonic Simulations models are within 10\% of each other, which we interpret as the systematic modeling uncertainty for the Stacked Model. This systematic error is still below our $15$\% statistical uncertainties on the mass, but with roughly eight more clusters they will become comparable. Looking ahead, if we want to use the full potential statistical power of the HSC survey or other comparable imaging surveys we need to reduce systematic modeling uncertainties.

Comparing the masses from the single-mass-bin fit and the Stacked Model method, we find that the single-mass-bin masses are systematically high by 3 to 7\%, depending on the profile model. We interpret this as a systematic bias from the single-mass-bin fitting technique that results from its lack of accounting for the mass and selection functions. Such a bias will become more important as samples of clusters with weak-lensing measurements increase.

\subsection{Mass Bias}
\label{sec:massbias}

\begin{figure*}[t]
\begin{center}
\includegraphics[width=6.5in]{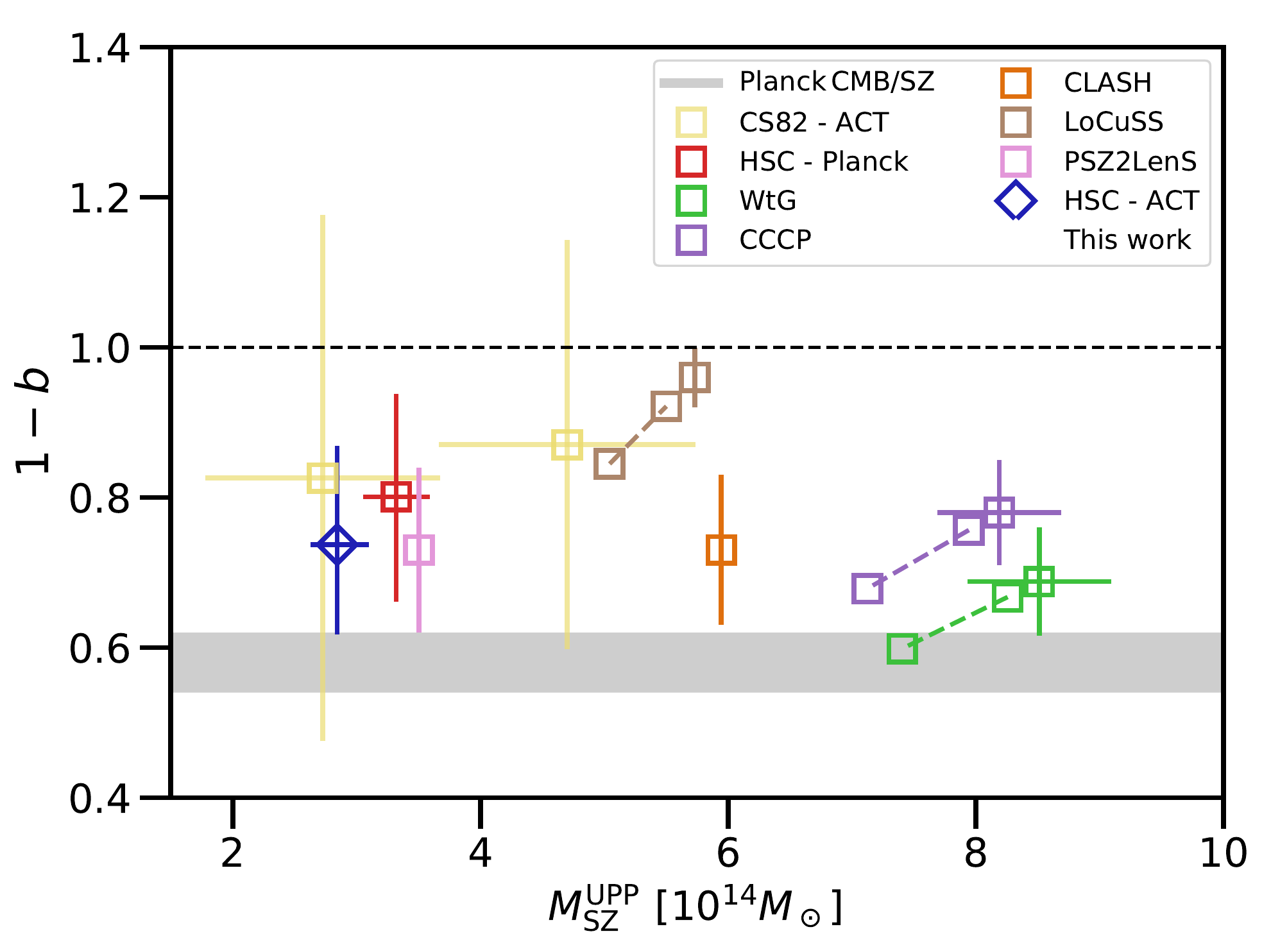}
\end{center}
\caption{Current comparison of $1-b$ for clusters with {\sl Planck} or ACT SZ masses. Here the data points show the ratios of $M^{\rm UPP}_{\rm SZ}$ to $M_{\rm WL}$, and we emphasize that these SZ masses are derived using the profile and X-ray mass scaling relation from \citet{Arnaud2010}, which assumes that the intracluster medium is in hydrostatic equilibrium. The grey band indicates the value of $1-b$ required to reconcile the \citet{PlnkSZCos2015} cluster cosmology results with the \citet{PlanckCMB2015} cosmological parameters from the primary CMB fluctuations. The result for this work is shown by the blue diamond from the Dark Emulator fit. Previous $1-b$ measurements by CS82-ACT \citep{Battaglia2016}, LoCuSS \citep{Smith2016}, CLASH \citep{Penna2017}, PSZ2LenS \citep{Sereno2017}, and HSC-{\sl Planck} \citep{Medez2017} are shown in yellow, brown, orange, pink ,and red squares, respectively. The green and purple squares with error bars show the original measurements from WtG \citep{WtG2014} and CCCP \citep{CCCP2015}, respectively, and the same colored squares connected by the dashed lines show the 3-15\% range for the Eddington corrected measurements calculated in \citet{Battaglia2016}. The weak-lensing mass calibration of ACT SZ masses by HSC is consistent with previous calibrations of ACT and {\sl Planck} SZ masses, including the LoCuSS measurements after the Eddington bias correction that was applied to \citet{WtG2014} and \citet{CCCP2015} is consistently applied to \citet{Smith2016}. The error bars here do not include sample variance (see Section~\ref{sec:massbias}).
}
\label{fig:1mbvMSZ}
\end{figure*}

We present the values for $1-b = \langle M_{\rm SZ} \rangle/\langle M_{\rm WL}\rangle$ in Table~\ref{tab:stacked_lensing_stacked_model_fit}. We estimate the average SZ mass using the lensing weight to be consistent with the stacking process in lensing measurement; 
$\langle M_{\rm SZ} \rangle = \sum_{ls} w_{ls} M_{{\rm SZ}, l}/\sum_{ls} w_{ls}=2.87^{+0.25}_{-0.20}\ \times 10^{14}\ {\rm M}_\odot$.
With only eight clusters the precision on the $1-b$ measurement we present in this work has comparable precision to previous measurements which have comparable or larger sample sizes. In Figure~\ref{fig:1mbvMSZ} we compiled from the literature previous measurements of $1-b$ on ACT and {\sl Planck} selected clusters from various weak-lensing measurements and include this measurement for comparison. In this work and previous work we directly compare ACT and {\sl Planck} SZ masses because both ACT and {\sl Planck} use the same SZ-mass scaling relations and pressure profiles \citep[for more details see][]{Battaglia2016}. However, in the recent ACTPol cluster catalog \citep{Hilton2017}, we found evidence of mass-dependent bias at the 2-$\sigma$ level when we compared the Planck and ACT SZ masses. We are ignoring that fact here since the mass dependence may only be the result of selection effects at the intersection of the {\sl Planck} and ACT samples. With a much larger sample from ACT in the near future we will be able to address this further.

Previous weak-lensing measurements of {\sl Planck} clusters \citet{WtG2014} and \citet{CCCP2015} show marginal evidence for a mass dependence in $1-b$. Combining these results with the previous ACT clusters result \citep{Battaglia2016} qualitatively strengthens this evidence. Since \citet{Battaglia2016} was published, several new measurements of $1 - b$ were made using {\sl Planck} SZ clusters. The combination of previous measurements with our new measurement indicates that the observational case for a mass dependence in $1-b$ is weaker (see Figure~\ref{fig:1mbvMSZ}). Simply fitting $1-b$ as a function of mass yields an exponent of $1-b \propto M^{-0.01 \pm 0.02}$. Here we have excluded the measurement from LoCuSS \citep{Smith2016} as their X-ray measurement using spectroscopic-like temperatures \citep{Martino14} gives $10\%$ higher X-ray masses in contrast to other measurements and they have a factor of two smaller errors than any other measurements of $1-b$. We caution against any strong conclusions from our simple analysis above, since a proper analysis requires compilation of all the {\sl Planck} and ACT SZ clusters with weak-lensing measurements, selecting and accounting for multiple weak-lensing measurements of these clusters, and precise Eddington bias corrections that account for the selection functions for each of these surveys convolved with the {\sl Planck} and ACT selection functions. Comparisons of $1-b$ across different measurements should include sample variance errors, but such errors are typically not included. We calculate the sample variance contribution to $\langle \Delta\Sigma \rangle (R) \rangle$ by randomly sampling halos from the simulations \citep{BBPSS} that satisfy the selection function and find this increases the diagonal of our covariance matrix by 20-30\%. Performing similar analyses on all the measurements in Figure~\ref{fig:1mbvMSZ} will strengthen our conclusions that currently these measurements are consistent and do not show evidence of a mass dependent in $1-b$.

\section{Discussion and Conclusions}
\label{sec:conclusion}
In this work we present weak-lensing observations from the HSC survey of eight ACTPol galaxy clusters selected by their SZ signal. The depth of the HSC survey allows us to make weak-lensing measurements of individual clusters out to $z=1.004$ with $S/N>2$. We stack the weak-lensing measurements of these eight clusters and employ a detailed model for the scaling relation, mass function, and selection function for this sample when we fit for the average weak-lensing mass. The combined signal-to-noise of the stacked weak-lensing measurement is $11.1$. We use three different mass profiles, the NFW profile (analytic), the Dark Emulator (N-body simulation), and the Baryonic Simulations (cosmological hydrodynamic simulations) to infer masses. The average weak-lensing masses inferred for this sample are $\langle M_{\rm WL} \rangle = 4.02^{+ 0.65}_{-0.61}\,10^{14}$M$_{\odot}$ , $3.89^{+ 0.61}_{-0.57}\,10^{14}$M$_{\odot}$  and $3.55^{+0.63}_{- 0.48}\,10^{14}$M$_{\odot}$ for the NFW, Dark Emulator, and Baryonic Simulations Models. We interpret the 10\% difference between Dark Emulator and the Baryonic Simulations models as a systematic modeling uncertainty, which is currently lower than our statistical uncertainty. We compare two methods for modeling the stacked signal and demonstrate that using a single mass-bin fit model will introduce systematic biases on the inferred average mass in the range of 3 to 7\%.

The weak-lensing measurement of $1-b$ from HSC for ACTPol selected clusters in this work is consistent with previous weak-lensing measurements from the CS82 survey for ACT selected clusters, although the latter has a factor of three times larger errors. Additionally, our measurement of $1-b$ is statistically consistent with previous measurements of {\sl Planck} SZ clusters and has comparable precision, with only eight clusters in the sample.

Direct comparisons to previous weak-lensing mass calibrations of SPT SZ masses are non-trivial. Unlike {\sl Planck} and ACT, SPT uses a different filter shape and scaling relation to infer SZ masses \citep[for details see ][]{Reic2013,Bleem2015}. In \citet{Hilton2017} we showed that re-calibrated ACT SZ masses from \citet{Hass2013} are in remarkable agreement with the SPT SZ masses in \citet{Bleem2015}. However, since there is no overlap in area between the \citet{Hilton2017} and the \citet{Bleem2015} samples, any comparisons would be indirect and require an understanding the sub-sample of selected clusters with weak-lensing measurements. The SPT collaboration is calibrating masses out to and beyond $z=1$ using the Hubble Space Telescope \citep{Schrabback2016} and SPT's calibrations are of comparable precision to our measurement \citep[see][]{Dietrich2017,Stern2018}. Future direct comparisons of $1-b$ between ACT and SPT will be possible as the area surveyed by ACT expands.

Additionally, as the area overlap between the HSC Survey and the ACT experiment increases over the next couple of years we expect the number of clusters to increase roughly proportional to the area. The final area overlap will be of order 1400 deg$^2$, which should roughly yield a factor of 40 times more clusters for the same depth CMB maps. This will dramatically reduce the errors on our mass calibration and we can address questions about the mass and redshift dependence of $1-b$.

The underlying goal of calibrating SZ masses is to infer cosmological parameters and we leave such analyses for future work. Here, we will qualitatively estimate how this measurement of $1-b$, when used as a prior for the {\sl Planck} cosmological cluster analyses, would translate into cosmological parameters. If we take the Dark Emulator fit for $1-b$, which falls between the $1-b$ values of $0.688 \pm 0.072$ \citep{WtG2014} and $0.780 \pm 0.092$ \citep{CCCP2015} used in the \citet{PlnkSZCos2015} cosmological analysis, then the inferred values of cosmological parameters like $\sigma_8$ and $\Omega_m$ would fall between the respective $\sigma_8$ and $\Omega_m$ values inferred from the \citet{WtG2014} and \citet{CCCP2015} $1-b$ priors. Our $1 - b$ measurement is not as precise as the \citet{WtG2014} and \citet{CCCP2015} priors so the resulting errors on $\sigma_8$ and $\Omega_\mathrm{m}$ if applied to the {\sl Planck} cluster data will also be larger. Thus, we currently cannot address whether there is any actual difference between SZ cluster abundances and CMB primary anisotropy measurements, which is illustrated by differences in $1-b$ (compare the measurements and grey band in Figure~\ref{fig:1mbvMSZ}). We expect that this difference will be revisited after revisions to the primary CMB constraints are made with current and future measurements of the optical depth from polarized primary CMB observations \citep[e.g.,][]{PlnkTau2016}, since the $1-b$ inferred here is degenerate with the optical depth for such analyses. We are excited that future HSC and ACT measurements present opportunities to further understand and characterize such differences soon.

\acknowledgments

The ACT project is supported by the U.S. National Science Foundation through awards AST-1440226, AST-0965625 and AST-0408698, as well as awards PHY-1214379 and PHY-0855887. Funding was also provided by Princeton University, the University of Pennsylvania, and a Canada Foundation for Innovation (CFI) award to UBC. ACT operates in the Parque Astron\'{o}mico Atacama in northern Chile under the auspices of the Comisi\'{o}n Nacional de Investigaci\'{o}n Cient\'{i}fica y Tecnol\'{o}gica de Chile (CONICYT). Computations were performed on the GPC supercomputer at the SciNet HPC Consortium and on the hippo cluster at the University of KwaZulu-Natal. SciNet is funded by the CFI under the auspices of Compute Canada,
the Government of Ontario, the Ontario Research Fund
– Research Excellence; and the University of Toronto.
The development of multichroic detectors and lenses
was supported by NASA grants NNX13AE56G and
NNX14AB58G.

The Hyper Suprime-Cam (HSC) collaboration includes the astronomical communities of Japan and Taiwan,
and Princeton University. The HSC instrumentation and software were developed by the National
Astronomical Observatory of Japan (NAOJ), the Kavli Institute for the Physics and Mathematics of the
Universe (Kavli IPMU), the University of Tokyo, the High Energy Accelerator Research Organization
(KEK), the Academia Sinica Institute for Astronomy and Astrophysics in Taiwan (ASIAA), and Princeton
University. Funding was contributed by the FIRST program from Japanese Cabinet Office, the Ministry
of Education, Culture, Sports, Science and Technology (MEXT), the Japan Society for the Promotion of
Science (JSPS), Japan Science and Technology Agency (JST), the Toray Science Foundation, NAOJ, Kavli
IPMU, KEK, ASIAA, and Princeton University.

This paper makes use of software developed for the Large Synoptic Survey Telescope. We thank the
LSST Project for making their code available as free software at  \texttt{http://dm.lsst.org}.

The Pan-STARRS1 Surveys (PS1) have been made possible through contributions of the Institute for
Astronomy, the University of Hawaii, the Pan-STARRS Project Office, the Max-{\sl Planck} Society and its
participating institutes, the Max {\sl Planck} Institute for Astronomy, Heidelberg and the Max Planck
Institute for Extraterrestrial Physics, Garching, The Johns Hopkins University, Durham University,
the University of Edinburgh, Queen's University Belfast, the Harvard-Smithsonian Center for
Astrophysics, the Las Cumbres Observatory Global Telescope Network Incorporated, the National
Central University of Taiwan, the Space Telescope Science Institute, the National Aeronautics and
Space Administration under Grant No. NNX08AR22G issued through the Planetary Science Division of the
NASA Science Mission Directorate, the National Science Foundation under Grant No. AST-1238877, the
University of Maryland, and Eotvos Lorand University (ELTE) and the Los Alamos National Laboratory.

Based on data collected at the Subaru Telescope and retrieved from the HSC data archive system,
which is operated by Subaru Telescope and Astronomy Data Center, National Astronomical Observatory
of Japan.

HM has been supported by the Jet Propulsion Laboratory, California Institute of Technology, under a contract with the National Aeronautics and Space Administration and Grant-in-Aid for Scientific Research from the JSPS Promotion of Science (No. 17H06600 and No. 18H04350). NB has been supported by Lyman Spitzer Jr. postdoctoral fellow and is currently supported by Simons Foundation. MT is supported by Grant-in-Aid for Scientific Research from the JSPS Promotion of Science (No. 15H03654, 15H05893, 15K21733, and 15H05892). TN acknowledges financial support from Japan Society for the Promotion of Science (JSPS) KAKENHI Grant Number 17K14273 and Japan Science and Technology Agency (JST) CREST Grant Number JPMJCR1414. KU acknowledges support from the Ministry of Science and Technology of Taiwan (grant MOST 103-2628-M-001-003-MY3) and from the Academia Sinica Investigator Award. EC acknowledges support from a STFC Ernest Rutherford Fellowship, grant reference ST/M004856/2. KM acknowledges support from the National Research Foundation of South Africa (grant number 93565). RD was supported by Anillo ACT-1417 and QUIMAL-160009. AvE was supported by the Vincent and Beatrice Tremaine Fellowship. LM is funded by CONICYT FONDECYT grant 3170846.

\bibliography{main}
\bibliographystyle{apj}

\appendix

\section{Details of covariance calculation}
\label{app:covariance}
\subsection{Covariance due to large scale structure}
We calculate the covariance due to large scale structure based on \cite{Oguri:2011}. For the $n$-th cluster at redshift $z_n$ we calculate the covariance due to the projection effect of large scale structure as
\begin{equation}
C^{\rm lss}_{ij,n} = \langle\Sigma_{{\rm cr}, ls}^{-1} \rangle^{-2} \int \frac{\ell d \ell}{2\pi} C^{\kappa\kappa}_\ell J_2\left(\frac{ \ell R_i}{\chi(z_n)}\right)J_2\left(\frac{\ell R_j}{\chi(z_n)}\right),
\end{equation}
where $J_2(x)$ is the second order Bessel function and $\chi(z)$ is the comoving distance at redshift $z$. We approximate the inverse critical surface mass density averaged over source galaxies $\langle\Sigma_{{\rm cr}, ls}^{-1} \rangle \sim \Sigma^{-1}_{\rm cr}(z_n, \langle z_s\rangle)$, where $\langle z_s\rangle$ is the mean redshift calculated using the photo-$z$ PDF stacked over source galaxies with the weak-lensing weight, i.e.,
\begin{equation}
\langle z_s \rangle = \frac{\int dz\ z \ P_{\rm stacked}(z)}{\int dz P_{\rm stacked}(z)},
\end{equation}
where $P_{\rm stacked}(z) = \sum_s \tilde{w}_{ls} P(z)/\sum_s \tilde{w}_{ls}$ and $s$ runs over source galaxies in all the radial bins after the color-color cut as a function of lens redshift (See Section \ref{sec:source_galaxy_selection} for details). The weak-lensing power spectrum $C^{\kappa\kappa}_\ell$ is defined as
\begin{equation}
C^{\kappa\kappa}_\ell = \int d\chi \frac{[W^{\kappa}(z)]^2}{\chi^2} P_m^{\rm NL}\left(k=\frac{\ell}{\chi};z\right),
\end{equation}
where $W^\kappa(z)$ is the lensing weight function defined by
\begin{equation}
W^\kappa(z)=\frac{\rho_m(z)\langle\Sigma_{\rm cr}^{-1}\rangle}{(1+z)}.
\end{equation}
Here we again use the same approximation $\langle\Sigma_{\rm cr}^{-1}\rangle\sim\Sigma_{\rm cr}^{-1}(z, \langle z_s \rangle)$. The non-linear matter power spectrum $P_m^{\rm NL}(k)$ is calculated by \textsc{CAMB} using the Halofit prescription \citep{Smith2003} with the fitting parameter derived in \cite{Takahashi2012}. We then calculate the covariance of the stacked lensing signal as
\begin{equation}
C^{\rm lss}_{ij} = \frac{\sum_n \tilde{v}_{n,i}\tilde{v}_{n,j}C^{\rm lss}_{ij,n}}{\sum_n \tilde{v}_{n,i} \sum_l\tilde{v}_{n,j}},
\end{equation}
where $\tilde{v}_{n,i}$ is the sum of the weak-lensing weight within the $i$-th bin of the $n$-th lens.
\subsection{Covariance due to intrinsic variations of projected cluster mass profile}
We estimate the covariance that accounts for intrinsic variations of projected cluster mass profile, which includes the scatter in concentration and the triaxiality effect, based on \cite{Umetsu2016}. They found that the intrinsic covariance of the lensing convergence $\kappa=\Sigma(R)/\Sigma_{\rm cr}$ can be well approximated by
\begin{equation}
\label{eq:Ckk}
C^{\kappa,{\rm int}}_{ij} = \alpha_{\rm int}^2 \kappa^2 \delta_{ij},
\end{equation}
where $\alpha_{\rm int}=0.2$ for $M_{200c}\sim10^{15}\msunh$ clusters. This formalism excludes the external contribution from $C^\mathrm{lss}$, which was formally included in $C^\mathrm{int}$ covariance by \citet{Gruen2015}. We convert Eq.~\ref{eq:Ckk} to the covariance of excess surface density $C^{\Delta\Sigma,{\rm int}}_{ij}$ , assuming the NFW profile with concentration-mass relation derived in \cite{Diemer2015}. Therefore, the covariance $C^{\Delta\Sigma,{\rm int}}_{ij}$ depends on cluster mass. Note that, however, covariances with difference cluster mass are actually similar in their shapes and have the same form once they are scaled by $r\rightarrow r/r_{\rm 200m}$. We then approximately calculate the covariance of the stacked lensing signal by
\begin{equation}
C_{ij}^{\rm int} = \frac{C^{\Delta\Sigma,{\rm int}}_{ij}(\langle M \rangle)}{N_{\rm cl}},
\end{equation}
where $\langle M_{\rm WL} \rangle$ (in $M_{\rm 500c}$) is the typical mass of our cluster sample and $N_{\rm cl}$ is the number of clusters. We vary the typical mass within $1.0 < \langle M_{\rm WL} \rangle/10^{14}M_\odot < 7.0$ and see how the $1-b$ constraints are affected. We find the change in $1-b$ is within 1\%, and thus use a fixed mass $\langle M_{\rm WL}\rangle = 3.5\times10^{14} M_\odot$ for the intrinsic covariance in the main text.
\section{Additional Systematic Tests}
\label{app:systematc_tests}
\subsection{Null Tests}
\subsubsection{B-mode Signal}
\begin{figure*}
\includegraphics[width=0.5\columnwidth]{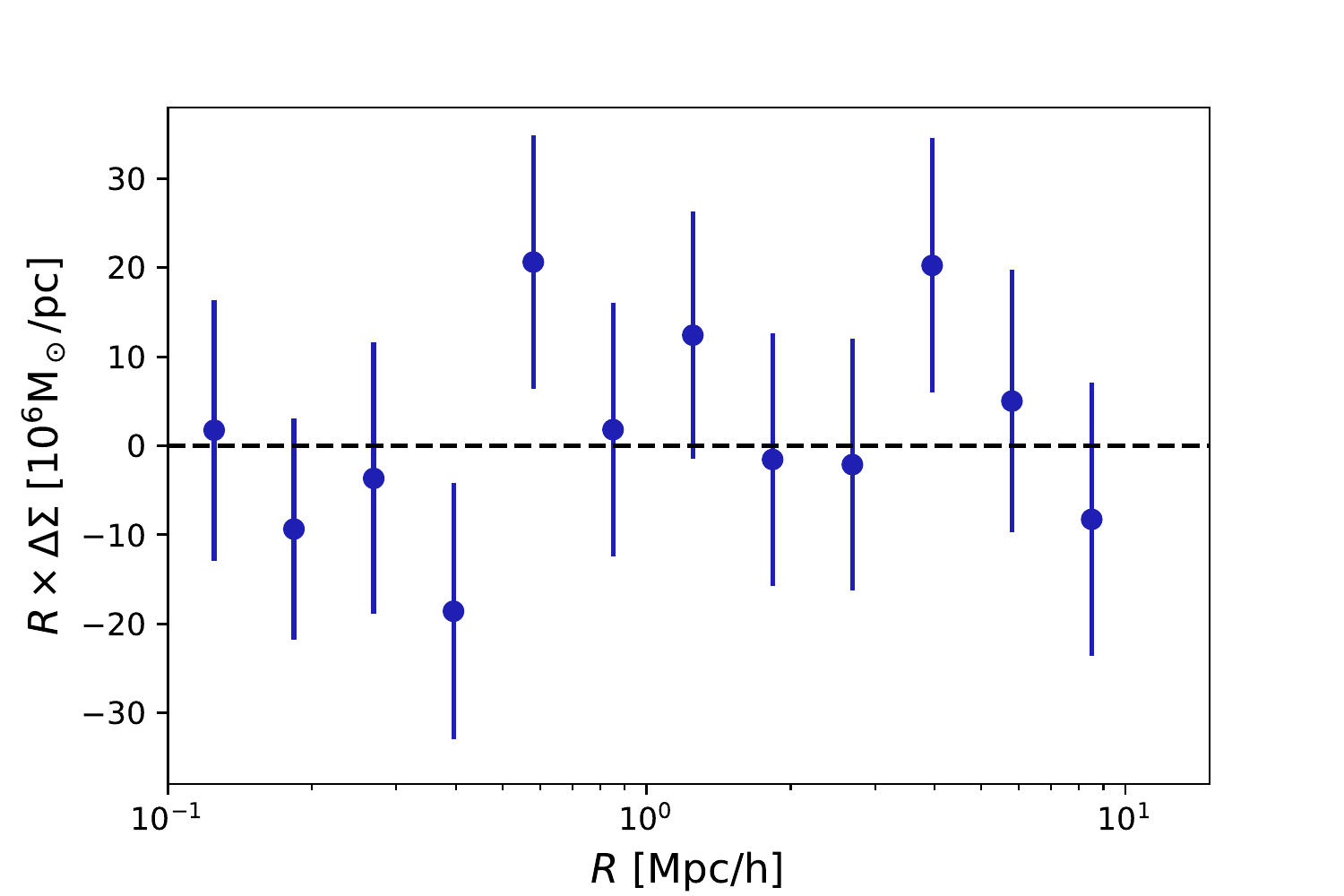}
\includegraphics[width=0.5\columnwidth]
{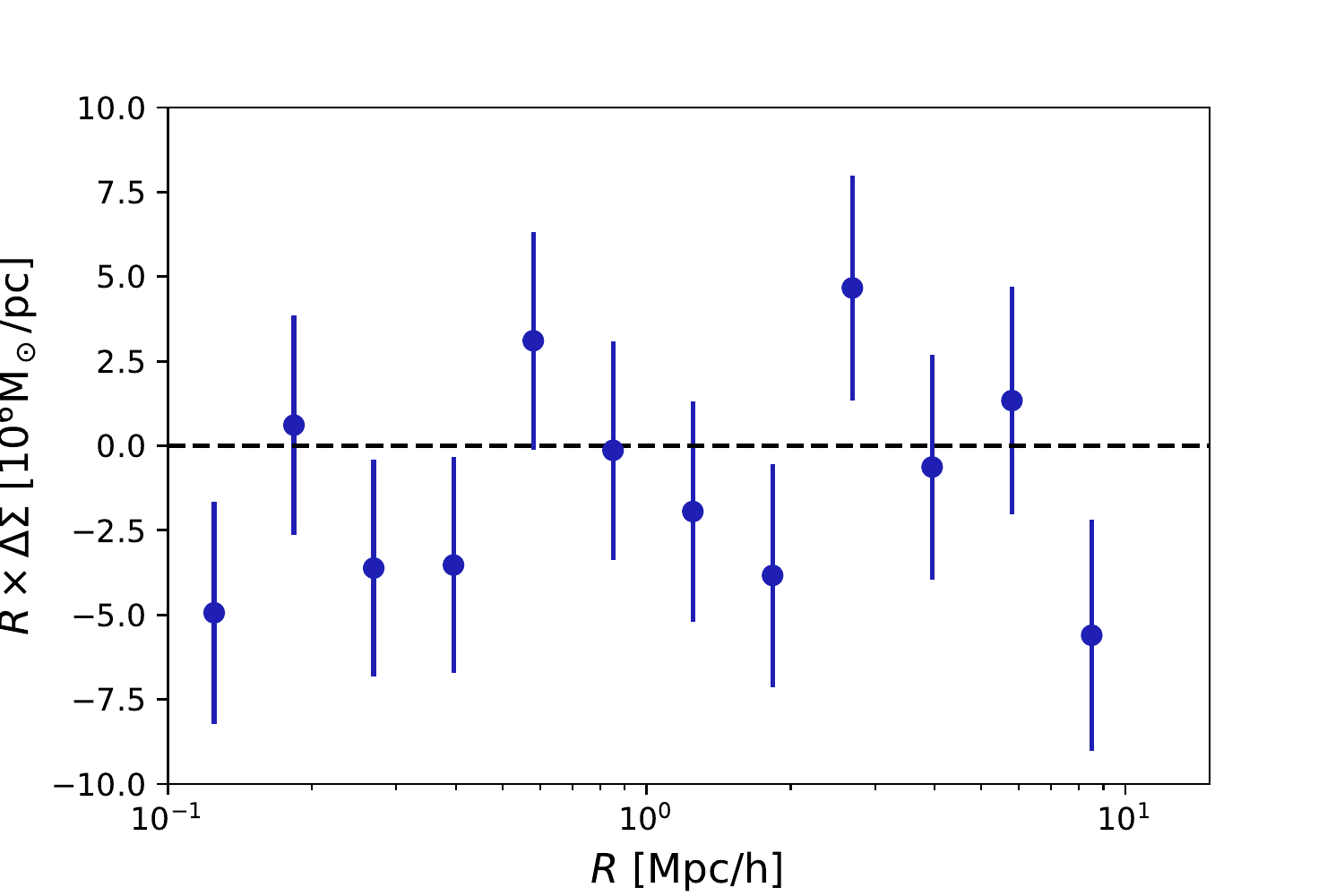}
\caption{{\it Left:} Stacked lensing B-mode signal around our ACTPol cluster sample. {\it Right:} Stacked lensing signal around 200 random points.}
\label{fig:B-mode}
\end{figure*}

Since weak lensing is caused by a scalar potential, a 45-degree rotated component from tangential shear, or B-mode, should be statistically consistent with zero. The left panel of Figure~\ref{fig:B-mode} shows the stacked B-mode signal around our ACTPol cluster sample. For our fitting range ($0.3\mpch<R<\mpch$), $\chi^2/dof=7.05/9$, and thus our B-model signal is consistent with zero.

\subsubsection{Random Signal}
If the PSF correction is imperfect, the lensing signal around random points will, statistically, be significantly different from zero. In this case, we need to correct for the imperfect PSF correction by subtracting the random signal from observed lensing signal. We generate random points within the HSC XMM field which follow the cluster redshift distribution based on the cluster mass function with the best fitting parameters of the NFW complete stacked fit (for details, see Section~\ref{section:complete_stacked_model}). The right panel of Figure~\ref{fig:B-mode} shows the stacked lensing signal around 200 random points. For our fitting range ($0.3\mpch<R<\mpch$), $\chi^2/dof=8.75/9$, which shows the random signal is consistent with zero.

\subsection{Lensing Signals with Different Photo-$z$ Methods}
\label{app:different_photozs}
We use MLZ photo-$z$ estimates for fiducial measurement as described in Section~\ref{sec:HSC_source_galaxies}. In this section we describe the details of other photo-$z$ methods in the HSC SSP catalog and check the consistency between ACTPol cluster lensing signals based on different photo-$z$ methods.

The HSC SSP catalog has photo-$z$ estimates based on the following methods; DEmP, Ephor, Franken-Z, Mizuki, and NNPZ \citep{Tanaka:2018}, in addition to MLZ. DEmP is a method designed to minimize major issues of conventional empirical methods, such as how to choose a proper fitting function and biases due to the population of a training data set, by introducing regional polynomial fitting and uniformly weighted training set \citep{Hsieh:2014}. Ephor is a neural network photo-$z$ code fed with de Vaucouleur flux and exponential flux. Franken-Z is a hybrid approach that combines the data-driven nature of machine learning and statistical rigor from template fitting. Mizuki is a Bayesian template fitting method which allows for simultaneously constraining physical properties of galaxies such as star formation and photo-$z$ \citep{Tanaka:2015}. NNPZ follows the method introduced by \cite{Cunha:2009}, a nearest neighbor method that finds nearest neighbors around an unknown object in the color/magnitude space from a reference sample and uses the reference redshift histogram as the PDF.

\begin{figure}
\begin{center}
\includegraphics[width=0.7\columnwidth]
{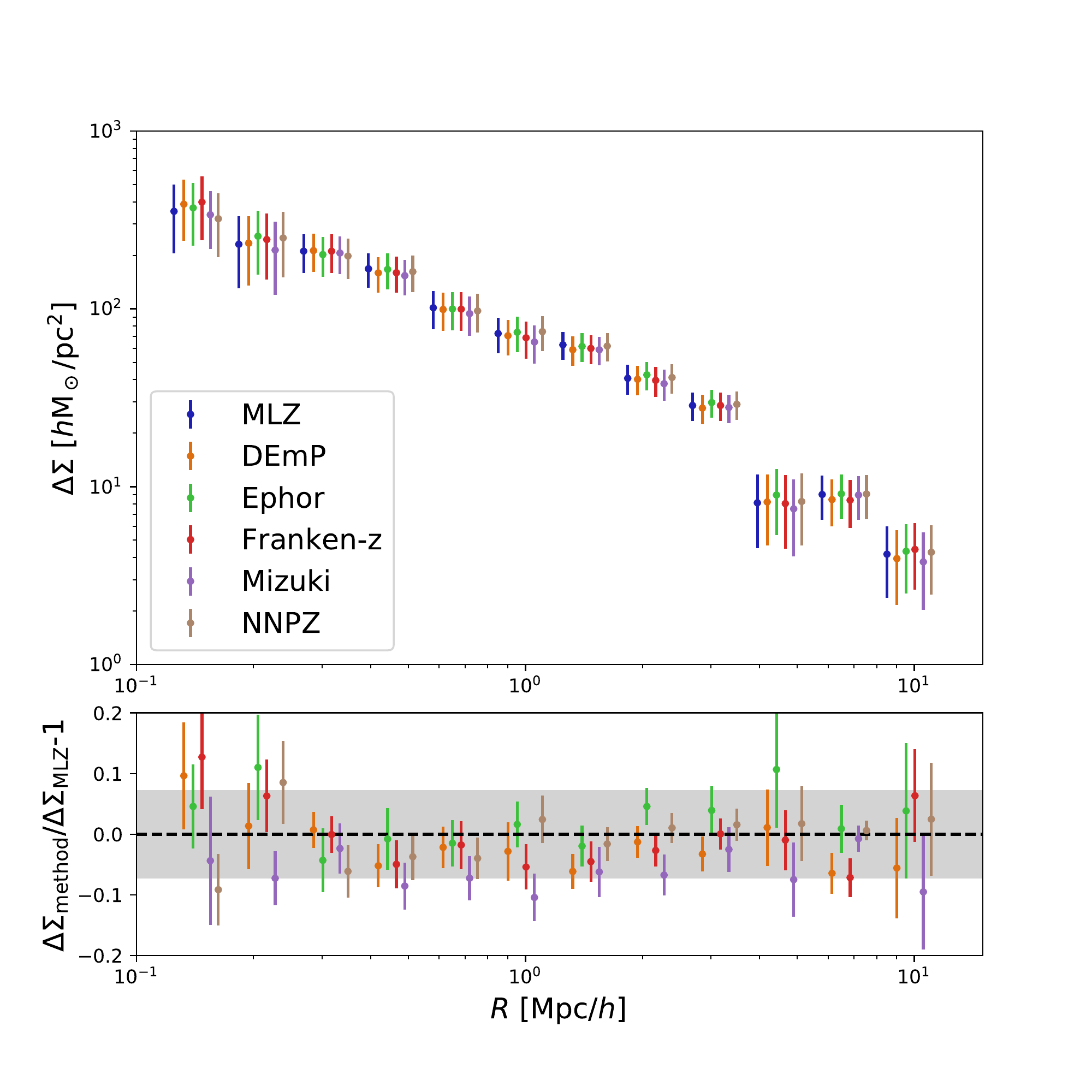}
\end{center}
\caption{{\it top:} The weak-lensing signals for different photo-$z$ methods. The data points with different photo-$z$ methods are shifted along x-axis for illustrative purposes. {\it bottom:} The fractional residuals between MLZ and other methods. Error bars are calculated by taking correlations between different lensing signals into account. The gray region shows the statistical error of our lensing measurement combined across the radial bins. See text for these details.}
\label{fig:lensing_different_photozs}
\end{figure}

The top panel of Fig.~\ref{fig:lensing_different_photozs} shows lensing signals measured with different photo-$z$s. The fractional residuals between MLZ and other photo-$z$ methods are shown in the bottom panel. When calculating error bars of fractional residuals, we account for the correlation between lensing signals based on different photo-$z$s. To estimate the correlation, we generate 18 realizations of galaxy shape catalogs with randomly rotated shapes. We then measure lensing signals around clusters in the same manner described in Section~\ref{sec:weak_lensing}, using different photo-$z$s. Although the signal itself has no tangential shear signal, we can still compute the correlation between lensing signals computed with different photo-$z$s. The inverse-variance weighted average of the fractional residual ranges from $-0.05\pm0.01$ (Mizuki) to $0.01\pm0.01$ (Ephor), which is smaller than the expected deviation due to statistical uncertainties of the stacked lensing signal, i.e., $\delta_{\Delta\Sigma}=\left[\sum_i\left(\Delta\Sigma(R_i)/\sigma_{\Delta\Sigma(R_i)}\right)^2\right]^{-1/2}$, where $\sigma_{\Delta\Sigma(R_i)}$ is the shape noise. Thus we conclude that the relative bias between photo-z methods are within the statistical uncertainties.

\subsection{Off-centering}
\label{off-centering}
If the cluster center used in the lensing measurement (in this case, the BCG position) is offset from the gravitational potential minimum, the lensing signal at inner radii around and below the scale of off-centering is diluted. Previous studies showed that the positions of BCGs are better tracers of potential minima than other optical tracers such as the luminosity-weighted centers \citep{Viola2015}, and than X-ray centers because of the large statistical uncertainties in the X-ray center \citep{George2012,WtG2014_2}. We expect this is also the case for our measurement. The ACT beam at 148 GHz is $1.4 \arcmin$ and with a 5$\sigma$ SZ detection we expect the astrometric uncertainties to be around $20\arcsec$ at $z \gtrsim 0.5$, which is similar to the typical off-centering except for the cluster J0229.6-0337.

\begin{figure}
\begin{center}
\includegraphics[width=0.6\columnwidth]{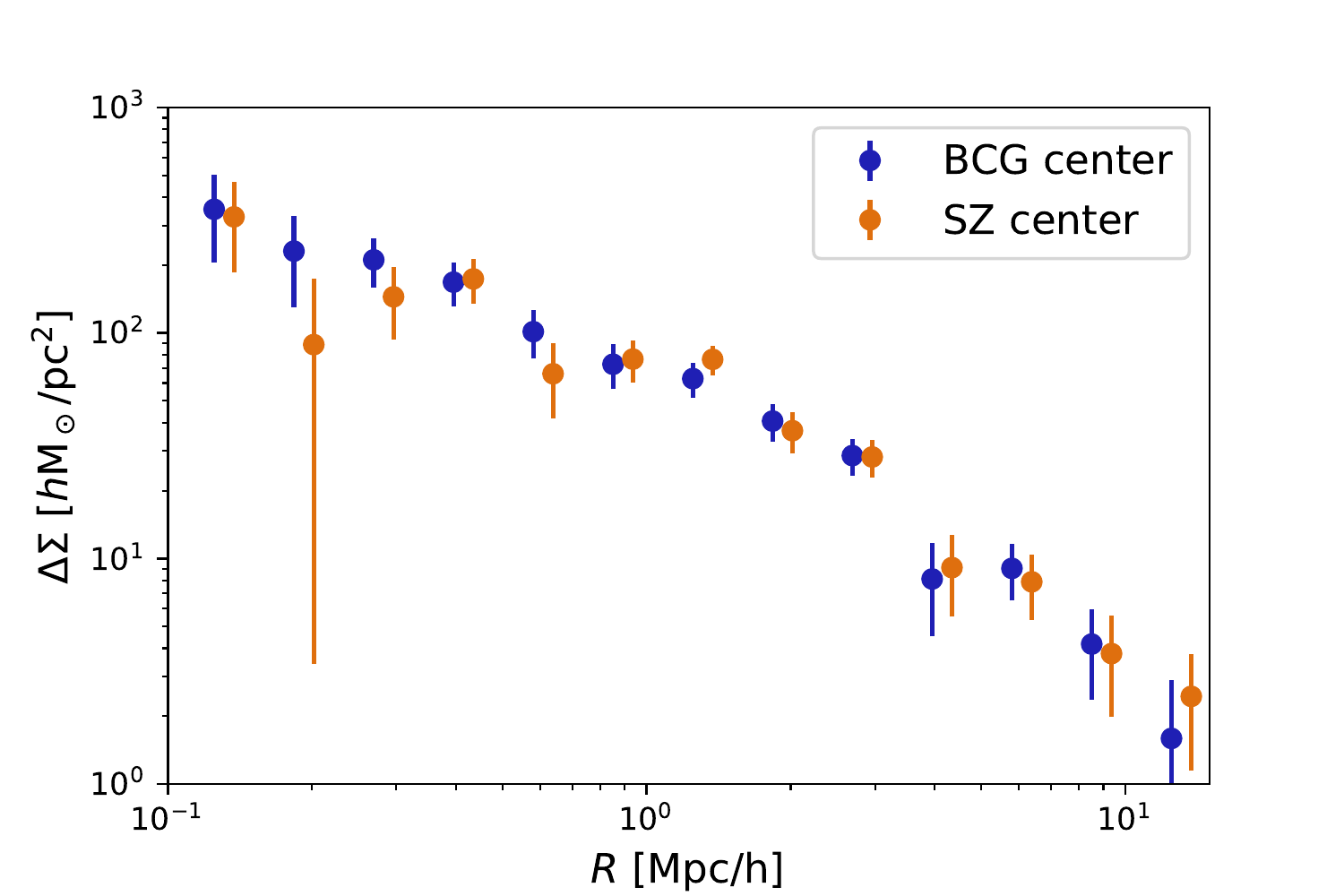}
\caption{The stacked weak-lensing signals calculated around different definitions of the cluster centers. The data points with for the SZ centers are shifted along x-axis for illustrative purposes. We do not see a significant difference between whether using the BCG center or the SZ center at the scales we use for model fitting ($R>0.3\mpch$). We use the BCG center for our fiducial measurement.}
\end{center}
\label{fig:lensing_offcentering}
\end{figure}

We check the impact of off-centering by comparing the lensing signal calculated with the BCG center to that with the SZ center. The comparison is shown in Figure~\ref{fig:lensing_offcentering}. We do not find a significant difference between these signals, especially at the scales we use when fitting the models ($R>0.3\mpch$), and thus it does not matter which center we use at this regime of signal-to-noise ratio for the stacked signal. Even if the lensing shear profile is affected by the off-centering effect, the enclosed mass should be properly extracted if we use the lensing signals to sufficiently large radii compared to the off-centering distance \citep[see][]{Oguri:2011}. For our fiducial measurement we use the BCG positions as centers. 

\section{Modeling Stacked Lensing Signal with Sparse Cluster Sampling}
In Section~\ref{section:complete_stacked_model}, we assumed in our model that the ACTPol clusters sample are representative of the underlying cluster distribution, which is a convolution of the halo mass function and ACTPol selection function. Thus, we integrated the lensing profile along redshift and SZ mass. Here we examine this assumption by replacing the integral in Eqs.~\ref{eq:delsigstack} and \ref{eq:mfivestack} with the summation of our cluster sample, i.e., 
\begin{equation}
\langle \Delta\Sigma \rangle (R_i) = \frac{1}{n_{\rm SZ}}\sum_j w_l(z_j) S(M_{{\rm SZ},j}, z_j) \int dM \frac{dn}{dM}(z_j) P(M_{{\rm SZ},j}|M)\Delta\Sigma(R_i,M,z_j),
\end{equation}
where $n_{\rm SZ}$ becomes
\begin{equation}
n_{\rm SZ} = \sum_j w_l(z_j) \int dM \frac{dn}{dM} (z_j) S(M_{{\rm SZ},j }, z_j)P(M_{{\rm SZ},j}|M),
\end{equation}
and $M_{\rm WL}$ becomes
\begin{equation}
\langle M_{\rm WL} \rangle = \frac{1}{n_{\rm SZ}}\sum_j w_l(z_j) S(M_{{\rm SZ},j}, z_j) \int dM \frac{dn}{dM}(z_j) P(M_{{\rm SZ},j}|M)M_{\rm WL}(M,z_j),
\end{equation}
where $M_{{\rm SZ},j}$ is the SZ mass before the Eddington bias correction. We then constrain $1-b$ in the same manner as Section~\ref{sec:massbias} for the NFW profile case. We find that the difference in the central value is less than one percent, which is negligible compared to statistical uncertainties. This demonstrates that our analysis method in Section~\ref{section:complete_stacked_model} is unbiased. We note that our test is not a general statement regarding such a bias and should not be applied to other cluster samples.

\end{document}